\documentclass[aps,prb,superscriptaddress,reprint,floatfix,nofootinbib]{revtex4-2}
\usepackage[final]{graphicx}
\usepackage{natbib}
\usepackage{hyperref}
\usepackage{amssymb}
\usepackage{amsmath}
\usepackage{mathtools}
\usepackage{mathrsfs}
\usepackage{xcolor}
\usepackage{bm}
\usepackage{multirow}
\usepackage{setspace}

\begin{document}
\title{Superexchange and spin-orbit coupling in monolayer and bilayer chromium trihalides}

\author{Kok Wee Song}
\email[Current affiliation email:]{k.song2@exeter.ac.uk}
\affiliation{Department of Physics and Astronomy, University of Exeter, Exeter, Devon EX4 4QL, UK}
\affiliation{National Graphene Institute, University of Manchester, Manchester M13 9PL, United Kingdom}
\author{Vladimir I Fal'ko}
\affiliation{National Graphene Institute, University of Manchester, Manchester M13 9PL, United Kingdom}
\affiliation{Henry Royce Institute for Advanced Materials, Manchester, M13 9PL, United Kingdom}
\affiliation{Department of Physics and Astronomy, University of Manchester, Oxford Road, Manchester, M13 9PL, UK}

\begin{abstract}
We build a microscopic model to study the intra- and inter-layer superexchange due to electrons hopping in chromium trihalides ($\mathrm{CrX}_3$, X= Cl, Br, and I). In evaluating the superexchange, we identify the relevant intermediate excitations in the hopping. In our study, we find that the intermediate hole-pairs excitations in the $p$-orbitals on X ion play a crucial role in mediating various types of exchange interactions. In particular, the inter-layer antiferromagnetic exchange may be realized by the hole-pair-mediated superexchange. Interestingly, we also find that these virtual hopping processes compete with each other leading to weak intra-layer ferromagnetic exchange. In addition, we also study the spin-orbit coupling effects on the superexchange and investigate the Dzyaloshinskii-Moriya interaction. Finally, we extract the microscopic model parameters from density functional theory for analyzing the exchange interactions in a monolayer $\mathrm{CrI}_3$.
\end{abstract}

\date{\today}

\maketitle
\section{introduction}

The realization of two-dimensional (2D) magnetism in an atomically-thin chromium trihalide ($\mathrm{CrX}_3$) is one of the recent breakthroughs in the field of 2D materials\cite{McGuire:ACScm27(2015),Burch:Nature563(2018),Gibertini:NatNano14(2019),Wang:AnnalPhys532(2020),Yao:Nanotech32(2021)}. Although 2D magnetism is relatively new to the field, the research has progressed at an incredible pace and demonstrated many intriguing magnetic phenomena due to low dimensionality\cite{Li:PRX10(2020),Lei:NanoLett21(2021),Wang:ACSNano16(2022),Arneth:PRB105(2022),Ghosh:PRB105(2022)}. Not to mention, this 2D magnet is a building block of van der Waals (vdW) heterostructures\cite{Geim:Nature499(2013),Novoselov:Science353(2016)} that can be integrated into devices to derive new functionality for spintronic applications\cite{Zhong:SciAdv3(2017),Cardoso:PRL121(2018),Zollner:PRB100(2019),Soriano:NanoLett20(2020),Rahman:ACSNano15(2021),Heissenbuettel:NanoLett21(2021)}. Despite the fascinating magnetic properties and potential applications, the understanding of the underlying physics of this low-dimensional magnetism is still incomplete and makes predicting the magnetic properties of these vdW materials challenging.

Stacking-dependent magnetism is perhaps one the of most intriguing properties in multilayer $\mathrm{CrX}_3$. In this regard, many experimental\cite{Huang:Nature546(2017),Huang:NatNanotech13(2018),Jiang:NatMat17(2018),Jiang:NatNanotech13(2018),Song:Science360(2018),Klein:Science360(2018),Wang:NatComm9(2018),Kim:NanotLett18(2018),Song:NanoLett19(2019),Chen:Science366(2019),Kim:PNAS116(2019),Klein:NatPhys15(2019),Li:NatMat18(2019),Guo:ACSNano15(2021)} and theoretical\cite{Sivadas:ACSnano18(2018),Jang:PRM3(2019),Jiang:PRB99(2019),Soriano:SSComm299(2019),Morell:2DMat6(2019),Gibertini:JPhysDapp54(2020),Sarkar:PRB103(2021),Wang:JPhysChemC125(2021),Xiao:PRRes3(2021),Yu:APLett119(2021)} efforts have been devoted to understanding the inter-layer exchange coupling and stimulating a lot of interest in Moir\'{e} magnetism\cite{Hejazi:PNAS117(2020),Li:PRB102(2020),Wang:PRL125(2020),Xie:NatPhys18(2021),Xu:NatNanoTech17(2021),Akram:NanoLett21(2021),Xiao:NanoLett22(2022),Shang:JPhysChemLett13(2022),Fumega:arXiv(2022)}. Nevertheless, modeling the inter-layer exchange in these materials quantitatively is a very challenging task\cite{Soriano:NanoLett20(2020)}, and the origin of such an inter-layer exchange is still an ongoing open problem.

\begin{figure}
\centering
\includegraphics[width=3.35in]{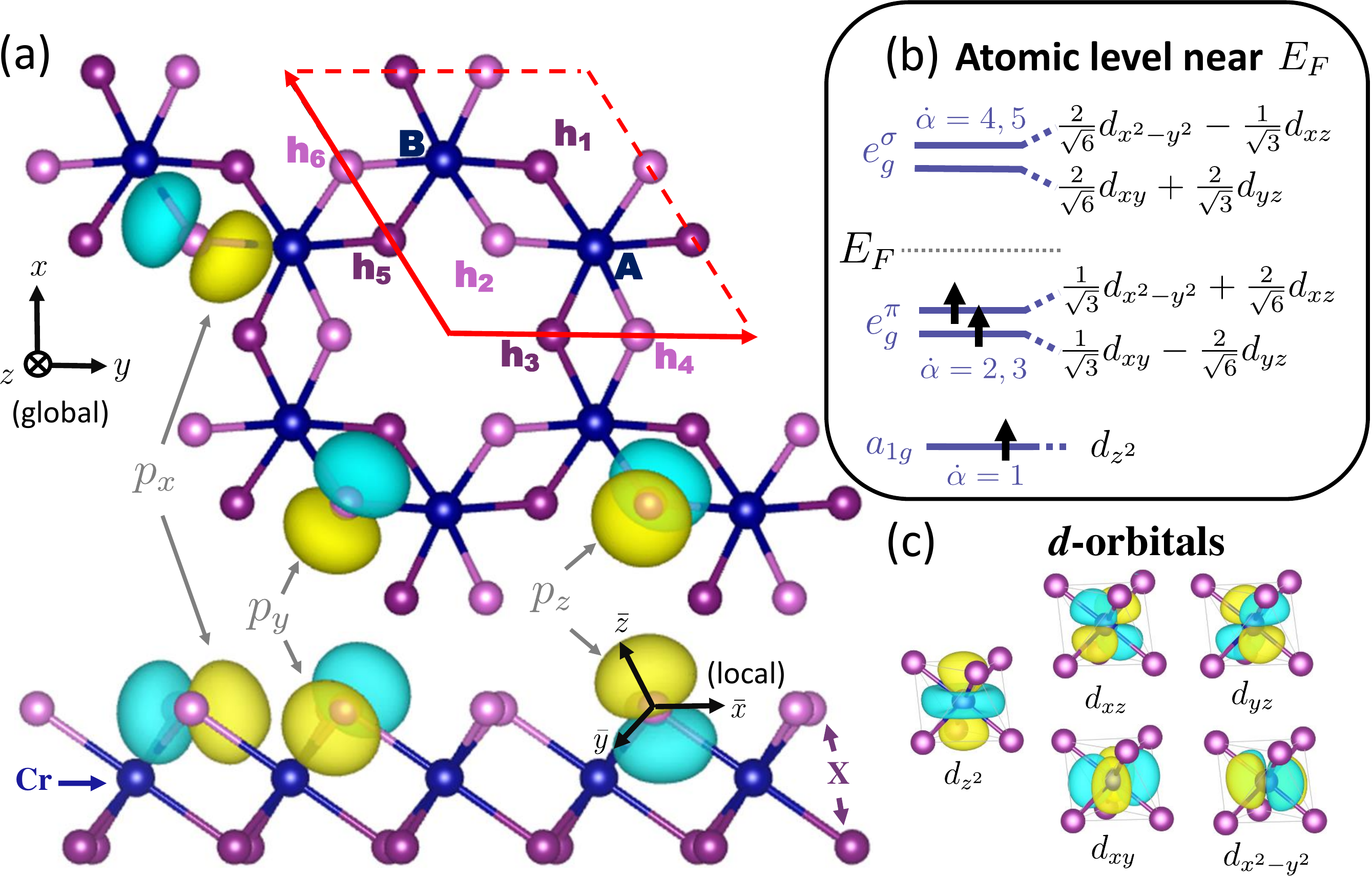}
\caption{(a) Top- and side-view of the crystal structure of $\mathrm{CrX}_3$. The dashed parallelogram is the unit cell of the crystal. The maximally localized Wannier functions (MLWF) of the X ion $p$ orbital are illustrated on the $h_2$ sublattice. The other $p$ orbital are related by the crystal symmetry. $\bar{x}\bar{y}\bar{z}$ is the local coordinate system for quantizing the angular momentum at $h_2$ which does not coincide with the $xyz$ global coordinate system. (b) the schematic atomic level with crystal field splitting near Fermi level $E_{F}$. (c) MLWF of the $d$-orbitals in the Cr atom for \textit{ab initio} TB model construction.
}\label{fig:1L-CrX3}
\end{figure}
In contrast to multilayer $\mathrm{CrX}_3$, the theory of intra-layer exchange in a monolayer $\mathrm{CrX}_3$ is very well established\cite{Zhang:JMatChemC3(2015),Lado:2DMat4(2017),Besbes:PRB99(2019),Torelli:2DMat6(2019),Wu:PhysChemChemPhys21(2019),Kashin:2DMat7(2020),Stavropoulos:PRR3(2021),Yadav:arXiv(2022)}, and the intra-layer ferromagnetic (FM) coupling can be intuitively explained by the superexchange due to the Cr-X-Cr hopping process. On the other hand, the inter-layer exchange coupling mediated by the Cr-X-X-Cr hopping process is relatively less well understood. The inter-layer superexchange may emerge as an ferromagnetic (FM) or antiferromagnetic (AFM) coupling depending on the stacking structure in a multilayer. This sensitive stacking-dependent behavior is a hallmark of competing effects of different exchange processes. For this reason, the inter-layer superexchange is inherently a complicated problem.

Besides, the spin-orbit coupling (SOC) effects in the superexchange are another outstanding theoretical problem in $\mathrm{CrX}_3$\cite{Soriano:NanoLett20(2020)}. Although these effects on the magnetic exchange may be weak, the SOC-mediated symmetric and antisymmetric exchanges have important implications for 2D materials. According to the Mermin-Wagner theorem, the long-range order in a low-dimensional system is susceptible to thermal fluctuations. To circumvent this in 2D magnets, the magnetic anisotropy that arises from the symmetric exchange is essential to stabilizing the magnetic order\cite{Lado:2DMat4(2017)}. 

Furthermore, unlike three-dimensional bulk, the Dzyaloshinskii-Moriya (DM) interaction due to the antisymmetric exchange may be enhanced significantly by the 2D materials' highly tunable electronic properties\cite{Liu:PRB97(2018),Liu:AIP8(2018),JaeschkeUbiergo:PRB103(2021),Ghosh:PhysicaB570(2019),Beck:JACSAu1(2021)}.  Interestingly, the enhanced DM interaction may lead to the emergence of chiral magnetic order under moir\'{e} engineering \cite{Akram:NanoLett21(2021)}. As suggested by experiments\cite{Chen:PRX8(2018),Chen:PRX11(2021),Mook:PRX11(2021),Cai:PRB104(2021)} and theories\cite{Kvashnin:PRB102(2020),JaeschkeUbiergo:PRB103(2021)}, the next-nearest-neighbor DM interaction may profoundly influence the magnetic ground state of $\mathrm{CrX}_3$ forming a gap at the Dirac points in the magnon dispersion. This may lead to a topological nontrivial ground state which hosts topological magnons at the edges\cite{McClarty:AnnRewCondMat13(2022)}. To explore these interesting magnetic phenomena, analyzing the competing effects due to hopping and spin-flip processes in the SOC-mediated superexchange also poses a challenge.

Therefore, we build a microscopic theory based on Anderson's superexchange approach\cite{Anderson:PR79(1950),Anderson:PR115(1959)} to tackle these challenges. In Sec. \ref{sec:model}, we will describe the problem and the model. In Sec. \ref{sec:GS}, we construct the theoretical framework for analyzing the ground state of the model. In Sec. \ref{sec:Cn}, we calculate the magnetic energy corrections for the ground state due to superexchange processes. We then verify our theory by applying it to $\mathrm{CrI}_3$ in Sec. \ref{sec:DFT}. This section also discusses how to extract our microscopic model parameters from the \textit{ab initio} calculations. Finally, we summarize the paper and discuss the outlook of this theoretical work in Sec. \ref{sec:conclusion}.

\section{Model Hamiltonian}\label{sec:model}

To study the magnetic ground state, we model the relevant electronic modes in a monolayer $\mathrm{CrX}_3$ by 
$
\mathcal{H}=\mathcal{H}_{E}+\mathcal{H}_{U}+\mathcal{H}'
$. The non-interacting Hamiltonian is  
\begin{subequations}
\begin{equation}\label{eqn:HE}
\mathcal{H}_E\!=\!\sum_{\mathbf{R}\alpha \sigma} \epsilon^{\alpha}_\mathbf{R}\psi^{\dagger}_{\mathbf{R}\alpha \sigma}\psi_{\mathbf{R}\alpha \sigma},
\end{equation}
where $\epsilon^{\alpha}_{\mathbf{R}}$  is onsite energy and
\begin{equation*}
\psi^{\dagger}_{\mathbf{R} \alpha \sigma}=
\begin{cases}
p^{\dagger}_{\mathbf{r}\alpha  \sigma},&\alpha=x,y,z\text{ at }\mathbf{R}=\mathbf{r}\\
d^{\dagger}_{\dot{\mathbf{r}}\dot{\alpha} \sigma},&\dot{\alpha}=1\dots5\text{ at }\mathbf{R}=\dot{\mathbf{r}}
\end{cases}
\end{equation*}
is the creation field operator at the lattice position $\mathbf{R}$ with orbital $\alpha$ and spin $\sigma$. Here, $p^{\dagger}_{\mathbf{r}\alpha \sigma}$ and $d^{\dagger}_{\dot{\mathbf{r}}\dot{\alpha} \sigma}$ are the creation field operators for $p$- and $d$-orbitals with $\mathbf{r}$ and $\dot{\mathbf{r}}$ being the X and Cr sublattices position. We introduce the dotted indices to explicitly distinguish the $d$-orbital from the $p$-orbital degrees of freedoms. The five $d$ orbitals split into $a_{1g}$, $e^\pi_{g}$, and $e^\sigma_{g}$ in the trigonal basis\cite{Georgescu:PRB105(2022)}, see Fig. \ref{fig:1L-CrX3}b. In our case, the $a_{1g}$ and $e^\pi_{g}$ are degenerate which combine together to form $t_{2g}$ ($\dot{\alpha}=1,2,3$) and $e^\sigma_{g}$ becomes $e_g$ ($\dot{\alpha}=4,5$) in the standard tetragonal basis. The three $p$ orbitals are labeled by $\alpha=x,y,z$. The interacting Hamiltonian is
\begin{equation}
\mathcal{H}_{U}\!=\!\sum_{\mathbf{R}\alpha \alpha'}\!\!\Big[U_{\mathbf{R}}^{\alpha \alpha'}\hat{n}_{\mathbf{R} \alpha}(\hat{n}_{\mathbf{R} \alpha'}\!-\!\tfrac{1}{2}\delta_{\alpha \alpha'}\!)\!-\!J_{\mathbf{R}}^{\alpha \alpha'}\hat{\mathbf{s}}_{\mathbf{R}\alpha}\!\cdot \hat{\mathbf{s}}_{\mathbf{R}\alpha'}\Big]\!,\label{eqn:HU}
\end{equation}
where $\hat{n}_{\mathbf{R}\alpha}=\frac{1}{2}\psi^{\dagger}_{\mathbf{R}\alpha \sigma}\psi_{\mathbf{R}\alpha \sigma}$ is  the occupation number operator, and $\hat{\mathbf{s}}_{\mathbf{R}\alpha}=\frac{1}{2}\psi^{\dagger}_{\mathbf{R}\alpha \sigma}\bm{\tau}_{\sigma \sigma'}\psi_{\mathbf{R}\alpha \sigma'}$ is the spin operator with $\bm{\tau}=(\tau^x, \tau^y,\tau^z)$ being the Pauli matrices. $U_{\mathbf{R}}^{\alpha \alpha'},J_{\mathbf{R}}^{\alpha \alpha'}>0$ are the onsite Hubbard and Hund interactions constants.

The tight-binding (TB) Hamiltonian is
$\mathcal{H}'=\mathcal{H}_t+\mathcal{H}^{\dagger}_t+\mathcal{H}_{d}+\mathcal{V}_\lambda$.
\begin{align}
\mathcal{H}_t=&\sum_{\dot{\mathbf{r}}\mathbf{r} }t_{\mathbf{r}\alpha ,\dot{\mathbf{r}}\dot{\alpha}}p^{\dagger}_{\mathbf{r}\alpha \sigma}d_{\dot{\mathbf{r}}\dot{\alpha} \sigma},\label{eqn:Ht}\\
\mathcal{H}_d=&\sum_{\dot{\mathbf{r}}\dot{\mathbf{r}}'}u_{\dot{\mathbf{r}}\dot{\alpha},\dot{\mathbf{r}}'\dot{\alpha}'}d^\dagger_{\dot{\mathbf{r}}'\dot{\alpha}' \sigma}d_{\dot{\mathbf{r}}\dot{\alpha} \sigma},\label{eqn:Hd}\\
\mathcal{V}_\lambda=&\sum_{\mathbf{r} }\sum_{\alpha_1 \alpha_2\alpha}i\lambda^{\alpha}\varepsilon_{\alpha\alpha_2 \alpha_1}\bar{\tau}^{\alpha}_{\bar{\sigma}_1 \bar{\sigma}_2} p^{\dagger}_{\mathbf{r}\alpha_1 \bar{\sigma}_1}p_{\mathbf{r}\alpha_2 \bar{\sigma}_2},\label{eqn:Vlambda}
\end{align}
\end{subequations}
where $t_{\mathbf{r}\alpha,\dot{\mathbf{r}}\dot{\alpha}}$ and $u_{\dot{\mathbf{r}}'\dot{\alpha}',\dot{\mathbf{r}}\dot{\alpha}}$ being the TB constants and $\lambda^\alpha$ being the spin-orbit coupling. $\varepsilon_{\alpha\alpha_2 \alpha_1 }$ is the levi-civita antisymmetric tensor. We stress that, in Eq. \eqref{eqn:Vlambda}, the spin in the Hamiltonian is quantized in the \emph{local} $\bar{x}\bar{y}\bar{z}$-system (Fig.\ref{fig:1L-CrX3}) where their axes coincide with the $p$-orbitals orientation. Hence, the barred spin index ($\bar{\sigma}$) and the barred Pauli's matrices ($\bar{\bm{\tau}}$) are differentiated from the unbarred $\sigma$ and $\bm{\tau}$ [in Eqs. \eqref{eqn:HE}- \eqref{eqn:Ht}] which spins are quantized in the \emph{global} $xyz$-coordinate system (Fig.\ref{fig:1L-CrX3}). The existence of such a preferential axis for spin is the result of symmetry broking due to spin-orbit coupling (more discussion in Appendix \ref{app:states}).

In this model, we include the Cr-Cr direct hopping, since the $d$-orbitals may not be highly localized, especially for X=I. This can be seen in the $\mathrm{CrI}_3$ quasiparticle dispersion that the valence band is not very flat\cite{MolinaSanchez:JMatChemC8(2020)}. Furthermore, we note that the X-X direct hopping between the $p$ orbitals may not be negligible as well. For instance, the hopping between the $h_1$ site and $h_2$ site may not be small due to the extended $p$ orbitals, see Fig. \ref{fig:1L-CrX3}. Nonetheless, we ignore the effect of X-X hopping in the paper, since they are irrelevant for the low-order perturbation calculations.

To analyze the ground state of the Hamiltonian, we can write down the grand partition function formally as $
\mathcal{Z}=\text{tr}\exp\left[-\beta\mathcal{H} \right]$
where $\beta =1/(k_BT)$\ with Boltzmann constant $k_B$ and temperature $T$. Treating $\mathcal{H}'$ as a perturbation, the partition function in the interacting picture reads as
\begin{equation}\label{eqn:Z}
\mathcal{Z}=\sum_{\Psi}\langle \Psi |\mathcal{T}e^{-\int_0^\beta d\tau \mathcal{H}'(\tau)}\mathrm{e}^{-\beta \mathcal{H}_0}|\Psi\rangle,
\end{equation}
where $\mathcal{T}$ is the time-order operator and $\tau$ is the imaginary time. $\mathcal{H}'(\tau)=\mathrm{e}^{\tau \mathcal{H}_0}\mathcal{H}'\mathrm{e}^{-\tau \mathcal{H}_0}$ with $ \mathcal{H}_0=\mathcal{H}_E+\mathcal{H}_U$. In Eq. \eqref{eqn:Z}, $|\Psi\rangle=\sum_{\{ n_{\mathbf{R}\alpha \sigma} \}}\Psi[\{ n^\alpha_{\mathbf{R} \sigma} \}]\prod_{\mathbf{R}\alpha \sigma}(\psi^{\dagger}_{\mathbf{R}\alpha\sigma})^{n_{\mathbf{R}\alpha \sigma}}|0\rangle$, where $|0\rangle$ being the vacuum with empty $p$ and $d$ orbitals. $n^\alpha_{\mathbf{R} \sigma}=0,1$ is the occupation number and $\{ n^\alpha_{\mathbf{R} \sigma} \}$ denotes a possible occupation-number configuration of a many-body state. $\Psi[\{n^\alpha_{\mathbf{R} \sigma} \}]$ is the many-body wavefunction amplitude in the occupation-number representation. The trace in the partition function \eqref{eqn:Z} is summed over all possible many-body wavefunctions $\Psi$ in the functional space.

To search for a stable ground state by exploring such a huge many-body space is a formidable task. Instead, a practical approach is to identify the low-energy subspace and focus on the analysis in the vicinity of this relevant space. In the next section, we discuss the construction of this low-energy subspace.

\section{Magnetic ground state} \label{sec:GS}

\begin{figure}
\includegraphics[width=3.3in]{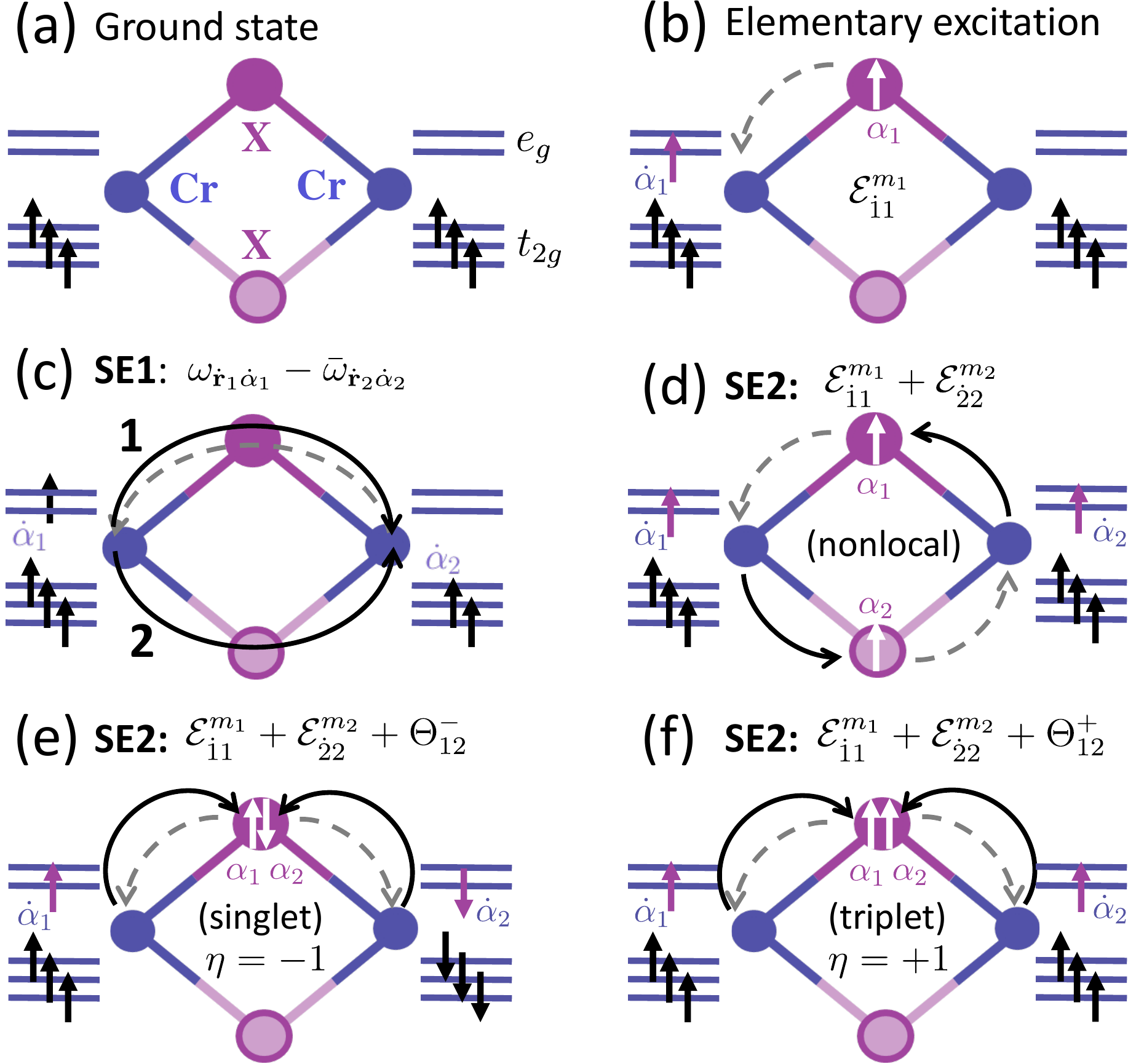}
\caption{Illustrations of the ground state and the excited states in the Mott phase. (a) Ground state: the electrons are localized at the Cr $d$ orbital giving rise to spin $\frac{3}{2}$. (b) Elementary excitation due to hopping: $\mathcal{E}^{m_1}_{\dot{1}1}$ is the excitation energy to create a $d$-electron (purple arrow) and a $p$-hole (white arrow). The dashed gray curve corresponds to the forward hopping process to create an excited state. (c) The excited states by moving an electron from one Cr to another Cr via X (SE1). The black curve (path 1 and path 2) is the backward hopping process that returns the excited state to the ground state. (d) - (f) The excitations that involve creating intermediate hole pair (SE2). The hole pair may be created nonlocally in (d) or locally in (e) and (f).}\label{fig:hopping}
\end{figure}

In the $\mathrm{CrX}_3$, the low-energy electronic modes are mostly composed of the $p$-orbitals on X ion and the $d$-orbitals on Cr ion. In particular, the $d$ orbitals are partially filled by an odd number of electrons. Based on the non-interacting analysis, the band structure of such an electronic system is a metal. However, the experimental and first-principle studies found that $\mathrm{CrX}_3$ is a semiconducting magnet with a wide band gap. This indicates the strongly-correlated nature of the magnetic ground state of $\mathrm{CrX}_3$. The magnetic phase of $\mathrm{CrX}_3$ is similar to the Mott insulating phase in the Hubbard model at half-filling. The charge carriers are highly localized in the insulating phase due to the strong Coulomb interactions. Indeed, this picture from the Mott physics is consistent to the magnetization measurement\cite{McGuire:ACScm27(2015)}, optical experiment \cite{Seyler:NatPhys14(2017)}, and the magnetization calculation in density functional theory (DFT) \cite{Zhang:JMatChemC3(2015),Lado:2DMat4(2017),Besbes:PRB99(2019),Torelli:2DMat6(2019),Wu:PhysChemChemPhys21(2019),Kashin:2DMat7(2020),Stavropoulos:PRR3(2021)}.

Motivated by these studies, we employ the approximation to project the whole many-body space $|\Psi\rangle$
onto the low-energy subspace $|\tilde{\Psi}\rangle$ which describes the Mott insulating state. This yields
\begin{equation}\label{eqn:Z_proj}
\mathcal{Z}\approx\mathcal{Z}_0\sum_{\tilde{\Psi}}\langle\tilde{\Psi}|\mathcal{T}e^{-\int_0^\beta d\tau \mathcal{H}'(\tau)}|\tilde{\Psi}\rangle,
\end{equation}
where the projection onto the Mott's state is
\begin{equation}\label{eqn:Mott}
|\Psi\rangle\approx|\tilde{\Psi}\rangle=
\prod_{\mathbf{r}\alpha \sigma}p^{\dagger}_{\mathbf{r}\alpha \sigma}\prod_{\dot{\mathbf{r}},\dot{\alpha}\leq3}\mathsf{d}^{\dagger}_{\dot{\mathbf{r}}\dot{\alpha}+}|0\rangle,
\end{equation}
and $\mathcal{Z}_0=\langle\tilde{\Psi}|\mathrm{e}^{-\beta \mathcal{H}_0}|\tilde{\Psi}\rangle$.
In the Mott's state, the $p_{x,y,z}$-orbitals are filled while the $d$-orbitals are half-filled in $\dot{\alpha}=1,2,3$ ($t_{2g}$) [see Fig.\ref{fig:1L-CrX3}(b)]. Because of the Hund's interaction, all the electrons' spins in the same Cr ion are deemed to parallel with the same unit vector, $\mathbf{s}_{\dot{\mathbf{r}}}=(s^x_{\dot{\mathbf{r}}},s^y_{\dot{\mathbf{r}}},s^z_{\dot{\mathbf{r}}})$. In Eq. \eqref{eqn:Mott}, $\mathsf{d}^{\dagger}_{\dot{\mathbf{r}}\dot{\alpha} \dot{\sigma}}=\chi^{\dot{\sigma}}_{\dot{\mathbf{r}}\sigma}d^{\dagger}_{\dot{\mathbf{r}}\sigma}$ ($\dot{\sigma}=\pm1$) is the field operator that creates a electron at $\dot{\mathbf{r}}$ with spin pointing in $\dot{\sigma}\mathbf{s}_{\dot{\mathbf{r}}}$  where $(\chi^+_{\dot{\mathbf{r}}\uparrow},\chi^+_{\dot{\mathbf{r}}\downarrow})=
[2(1+s^z_{\dot{\mathbf{r}}})]^{-\frac{1}{2}}(1\!+\!s^z_{\dot{\mathbf{r}}},\;s^x_{\dot{\mathbf{r}}}\!+is^y_{\dot{\mathbf{r}}})$ is the spin wave function  ($\chi^-_{\dot{\mathbf{r}}\sigma}=\sum_{\sigma'}i\tau^y_{\sigma \sigma'}\bar{\chi}^+_{\dot{\mathbf{r}},\sigma'}$).

As compared to Eq. \eqref{eqn:Z}, we note that the projected ground state in Eq.\eqref{eqn:Z_proj} has a simple many-body wave function. It is essentially the product of each local spin wave function, $\tilde{\Psi}=\prod_{\dot{\mathbf{r}},\dot{\alpha}\leq3}\chi^+_{\dot{\mathbf{r}}\sigma}$. Therefore, the trace over $\Psi$ in Eq. \eqref{eqn:Z} reduces to the trace over the many-body state with all possible local spin orientations with amplitude $\tilde{\Psi}$ in Eq. \eqref{eqn:Z_proj}. We mention that this many-body subspace is still very large and degenerates in $\mathcal{H}_0$. As we can see, without the hopping ($\mathcal{H}'=0$, atomic limit), each of the different spin configurations with arbitrarily local spin orientations has the same energy, since $\mathcal{H}_0$ is merely determined by the occupation number. But, this degeneracy will be lifted by the electrons hopping process. As a result, certain spin configurations become more energetically favorable than the others. This ultimately leads to the magnetic order in the ground state.

To investigate these hopping effects, we calculate the expectation value in Eq.\eqref{eqn:Z_proj} perturbatively by using cumulant expansion\cite{Shankar:RMP66(1994)}
\begin{align}\label{eqn:connect}
\langle\tilde{\Psi}|
&\mathcal{T}e^{-\int_0^\beta d\tau \mathcal{H}_t(\tau)}|\tilde{\Psi}\rangle\!= \!\exp \left(\sum_{n=0}^{\infty} \! \frac{C_n}{n!}\right),
\end{align}
where the \textit{connected} $n$-correlation functions ($n$-order cumulants) are
\begin{align}\label{eqn:Gc}
C_2&=\mu_2,\quad C_4=\mu_4-3 \mu_2^2, \quad C_5= \mu_5, \notag\\
C_6&=\mu_6-15 \mu_4\mu_2+30\mu_2^3,
\end{align}
with $\mu_n=\langle\tilde{\Psi}| \mathcal{T}[-\int^\beta_0 d \tau \mathcal{H}'(\tau)]^n|\tilde{\Psi}\rangle$ ($n$-th moment). We note that $\mu_3=\mu_1=0$ (Appendix \ref{app:link-cluster}). In this paper, we only consider the correlation function $C_n$ up to $n=6$.

So far, we only discussed the theoretical formalism of the problem. In the next section, we will focus on the calculation of these correlation functions $C_n$. These $C_n$ will give the exchange energies corrections' to the ground state. Consequently, the $\mathbf{s}_{\dot{\mathbf{r}}}$-dependent energy corrections appear in Eq. \eqref{eqn:connect} which yield the effective spin model with various exchange couplings.

\section{Exchange energies and $C_n$}\label{sec:Cn}

The most important information for evaluating $C_n$ is the spectrum of the unperturbed Hamiltonian $\mathcal{H}_0$. The excited states in the spectrum serve as the intermediate states in the perturbation calculations. Hence, in the first subsection, we will focus on solving the eigenvalue problem of $\mathcal{H}_0$. We also note that the full spectrum of $\mathcal{H}_0$ is not required for obtaining an accurate result. Therefore, in our paper, we only consider the relevant low-energy excitations as follows: (1) one-electron excitation on Cr's $d$-orbital and (2) one- and two-hole excitations on X's $p$-orbital. We will use these low-energy excitations to evaluate $C_n$ in the next subsection. 

\subsection{Quasiparticle spectrum in atomic limit}\label{sec:ExStates}

\begin{table}
\caption{\label{tbl:eigenenergy} The low-energy excited eigenstates and energies of $\mathcal{H}_0$. The excitation energies $E_{\mathbf{r}\alpha}$ and $E_{\dot{\mathbf{r}}\dot{\alpha}}$ are given in Eq. \eqref{eqn:1h-energy}. The two-hole state correlation energy $\Theta^{\eta,\alpha \alpha'}_{\mathbf{r}\mathbf{r}'}$ can be found in Eq. \eqref{eqn:2-h_E}. $\lambda^m_{\dot{\alpha}}$ and $v_{k,\dot{\alpha}}^m$ are obtained by solving Eq. \eqref{eqn:v-lambda} where $m=0,\dots3$ is the to label of the eigenstates.}
\begin{ruledtabular}
\begin{tabular}{lll}
 &$\mathcal{H}_0$ eigenstates  & eigenenergies\\
\hline
\multirow{2}{*}{$p_{x,y,z}$} & $p_{\mathbf{r}\alpha\sigma}|\tilde{\Psi}\rangle$  & $\bar{\nu}_{\mathbf{r}\alpha}=-E_{\mathbf{r}\alpha}$
\\
  & $\Phi^{\eta,\alpha\alpha'}_{\mathbf{r}\sigma\sigma'}|\tilde{\Psi}\rangle$ & $-E_{\mathbf{r}\alpha}-E_{\mathbf{r}\alpha'}+\Theta^{\eta,\alpha \alpha'}_{\mathbf{r}\mathbf{r}'}$
\\
\hline
 \multirow{2}{*}{$a_{1g},$}  & $\mathsf{d}_{\dot{\mathbf{r}} \dot{\alpha}+}|\tilde{\Psi}\rangle$  & $\bar{\omega}_{\dot{\mathbf{r}}\dot{\alpha}}=-E_{\dot{\mathbf{r}}\dot{\alpha}}$
\\
 $e^{\pi}_{g}$ &$\mathsf{d}^\dagger_{\dot{\mathbf{r}}\dot{\alpha}-}|\tilde{\Psi}\rangle$ & $ \omega_{\dot{\mathbf{r}}\dot{\alpha}}= E_{\dot{\mathbf{r}}\dot{\alpha}}\!+\!\!\displaystyle\sum_{ \dot{\alpha}'\neq  \dot{\alpha}}^{3}J^{ \dot{\alpha} \dot{\alpha}'}_{\dot{\mathbf{r}}}\!+\!\tfrac{1}{2}U^{\dot{\alpha}\dot{\alpha}}_{\dot{\mathbf{r}}}$
\\
\hline
\multirow{2}{*}{$e^{\sigma}_{g}$}  & $\mathsf{d}^\dagger_{\dot{\mathbf{r}}\dot{\alpha}+}|\tilde{\Psi}\rangle$ 
& $\omega_{\dot{\mathbf{r}}\dot{\alpha}}=E_{\dot{\mathbf{r}}\dot{\alpha}}$
\\
  & $\displaystyle\sum_{k=0}^3v^m_{k,\dot{\alpha}}\phi^{k,\dagger}_{ \dot{\mathbf{r}} \dot{\alpha}}|\tilde{\Psi}\rangle$ 
&
 $ \varpi^m_{\dot{\mathbf{r}}\dot{\alpha}}=E_{\dot{\mathbf{r}}\dot{\alpha}}+\displaystyle\sum_{ \dot{\alpha}'\neq  \dot{\alpha}}^{3}J^{ \dot{\alpha} \dot{\alpha}'}_{\dot{\mathbf{r}}}+\lambda^m_{\dot{\alpha}}$
\end{tabular}
\end{ruledtabular}
\end{table}

In the atomic limit, the problem reduces to a single site problem, since we can construct all the many-body states by knowing only the single site properties. The single-site physics is governed by the quasiparticle excitation which is a creation of electron (above Fermi level) or hole (below Fermi level).

\textbf{One-hole}--- The excitation energy of one-hole is straightforward to evaluate
\begin{equation}\label{eqn:1-h_E}
\mathcal{H}_0\begin{bmatrix}
p_{\mathbf{r}\alpha \sigma}\\
\mathsf{d}_{\dot{\mathbf{r}}\dot{\alpha}\dot{\sigma}}
\end{bmatrix}|\tilde{\Psi}\rangle=-\begin{bmatrix}
E_{\mathbf{r}\alpha}p_{\mathbf{r}\alpha \sigma}\\
E_{\dot{\mathbf{r}}\dot{\alpha}}\mathsf{d}_{\dot{\mathbf{r}}\dot{\alpha}\dot{\sigma}}
\end{bmatrix}|\tilde{\Psi}\rangle,
\end{equation}
where the one-hole energies are
\begin{equation}
\begin{bmatrix}
E_{\mathbf{r}\alpha}\\
E_{\dot{\mathbf{r}}\dot{\alpha}}
\end{bmatrix}=
\begin{bmatrix}
\epsilon_{\mathbf{r}}^{\alpha}+\sum_{\alpha'\neq\alpha}U^{\alpha\alpha'}_{\mathbf{r}}-\frac{1}{2}U^{\alpha \alpha}_{\mathbf{r}}\\
\epsilon_{\dot{\mathbf{r}}}^{\dot{\alpha}}+\frac{1}{2}\sum_{\dot{\alpha}'\neq\dot{\alpha}}^3(U^{\dot{\alpha}\dot{\alpha}'}_{\dot{\mathbf{r}}}-J^{\dot{\alpha}\dot{\alpha}'}_{\dot{\mathbf{r}}})
\end{bmatrix}
\label{eqn:1h-energy}
\end{equation}
To calculate the above, we note that the eigenvalues of the Hubbard-$U$ interacting Hamiltonian is obtained by counting the number of the electrons and holes in the state. To calculate the eigenvalues of Hund interacting Hamiltonian, it is very instructive to write the scalar product of spin operators as
\begin{equation}\label{eqn:s1.s2}
\hat{\mathbf{s}}_{\mathbf{R}\alpha}\cdot\hat{\mathbf{s}}_{\mathbf{R}\alpha'}=\frac{1}{2}[(\hat{\mathbf{s}}_{\mathbf{R}\alpha}+\hat{\mathbf{s}}_{\mathbf{R}\alpha'})^2-(\hat{\mathbf{s}}_{\mathbf{R}\alpha'})^2-(\hat{\mathbf{s}}_{\mathbf{R}\alpha'})^2].
\end{equation} 
Therefore, the Hund interacting energies can be obtained by rewriting the electrons or holes pair into total angular momentum (singlet/triplet) representation. Also, we note that all the excitation energies are measured from the Mott's ground state. Namely, we set $\mathcal{H}_0|\tilde{\Psi}\rangle\equiv0$.

\textbf{Two-hole}--- The spectrum for creating a two-hole state may be complicated due to the correlation effects. If the two created holes sitting on different lattice (nonlocal hole-pair), the excitation energy is simply the addition of the one-hole energy [Eq.\eqref{eqn:1h-energy}]. Once the two holes are created on the same lattice (local hole-pair), we need to take into account the effect of interactions. To calculate this excitation, we rewrite the two-hole creation field operator into spin-singlet ($\eta=-1$) and spin-triplet ($\eta=+1$) representation as 
\begin{equation}
\Phi^{\eta,\alpha\alpha'}_{\mathbf{R}\mathbf{R}'\sigma\sigma'}=\frac{1}{2}(\psi_{\mathbf{R}\alpha \sigma}\psi_{\mathbf{R}'\alpha' \sigma'}+\eta \psi_{\mathbf{R}\alpha \sigma'}\psi_{\mathbf{R}'\alpha' \sigma}).    
\end{equation}
Using Eq. \eqref{eqn:s1.s2}, this yields
\begin{equation}\label{eqn:2-h_E}
\mathcal{H}_0\Phi^{\eta,\alpha \alpha'}_{\mathbf{R}\mathbf{R}'\sigma \sigma'}|\tilde{\Psi}\rangle\!=\!(\Theta^{\eta,\alpha \alpha'}_{\mathbf{R}\mathbf{R}'}\!-E_{\mathbf{r}\alpha}-E_{\mathbf{r}\alpha'})\Phi^{\eta,\alpha \alpha'}_{\mathbf{R}\mathbf{R}',\sigma \sigma'}|\tilde{\Psi}\rangle,
\end{equation}
where $\Theta^{\eta,\alpha \alpha'}_{_{\mathbf{R}\mathbf{R}'}}=\frac{1}{2}[U^{\alpha\alpha'}_{\mathbf{R}}-(2\eta+1)(1-\delta_{\alpha \alpha'})J^{\alpha \alpha'}_{\mathbf{R}}]\delta_{\mathbf{R}\mathbf{R}'}$ is the correlation energy coming from the interaction between two holes. In this paper, we do not consider the local hole-pairs in the $d$-orbital since their effects only emerges in the higher-order perturbation calculations.

\textbf{One-electron}---
We only focus on the electron's excitation spectrum on the $d$ orbitals since the $p$ orbitals are filled in the Mott states. Furthermore, we ignore the two or more electron excitations on the same lattice by assuming strong Hubbard-$U$ interactions. This is very likely to be the case for $\mathrm{CrX}_3$. Therefore, we limit ourselves to calculating the one-electron excitations (the spectrum of four-particle states on a one-Cr site). We begin with the excitations with $\dot{\alpha}\leq3$. This excitation is the eigenstates of $\mathcal{H}_0$
\begin{equation}
\mathcal{H}_0\mathsf{d}^\dagger_{\dot{\mathbf{r}}\dot{\alpha}\dot{\sigma}}|\tilde{\Psi}\rangle\!
=\!
\delta_{\dot{\sigma},-}(E^{\dot{\alpha}}_{\mathbf{r}}+\tfrac{1}{2}U^{\alpha \alpha}_{\dot{\mathbf{r}}}\!+\!
\sum_{\dot{\alpha}'\neq\dot{\alpha}}^3J^{\dot{\alpha}\dot{\alpha}'}_{\dot{\mathbf{r}}})\mathsf{d}_{\dot{\mathbf{r}}\dot{\alpha}\dot{\sigma}}|\tilde{\Psi}\rangle.
\end{equation}
For $\dot{\alpha}>3$, it is straightforward to show that
\begin{align}
\mathcal{H}_0&\mathsf{d}^\dagger_{\dot{\mathbf{r}}\dot{\alpha}\dot{\sigma}}|\tilde{\Psi}\rangle
\!=\!
\Big\{
\Big[\epsilon_{\dot{\mathbf{r}}\dot{\alpha}}\!+\!\frac{1}{2}\sum_{\dot{\alpha}'=1}^3(
U_{\dot{\mathbf{r}}}^{\dot{\alpha}\dot{\alpha}'}\!-\!J_{\dot{\mathbf{r}}}^{\dot{\alpha}\dot{\alpha}'}\delta_{\dot{\sigma},+})\Big]\mathsf{d}^\dagger_{\dot{\mathbf{r}}\dot{\alpha}\dot{\sigma}}\notag\\
&
-\delta_{\dot{\sigma},-}\sum_{\dot{\alpha}'=1}^3\sum_{\eta=\pm1}\tfrac{2\eta-1}{2} J_{\dot{\mathbf{r}}}^{\dot{\alpha}\dot{\alpha}'}\Phi^{\eta,\dot{\alpha}\dot{\alpha}',\dagger}_{\dot{\mathbf{r}},-,+}\mathsf{d}_{\dot{\mathbf{r}}\dot{\alpha}',+}\Big\}|\tilde{\Psi}\rangle,\label{eqn:H0d-}
\end{align}
where, in the last line, $\Phi^{\eta,\dot{\alpha}\dot{\alpha}',\dagger}_{\dot{\mathbf{r}},-+}\mathsf{d}_{\dot{\mathbf{r}}\dot{\alpha}',+}$ is the operator that creates a spin-singlet ($\eta=-1$) or spin-triplet ($\eta=+1$) electron-pair in $\dot{\alpha}$ and $\dot{\alpha}'$ orbital. As we can see in the first line of Eq. \eqref{eqn:H0d-} that, $\mathsf{d}^\dagger_{\dot{\mathbf{r}}\dot{\alpha},+}|\tilde{\Psi}\rangle$ is the eigenstate of $\mathcal{H}_0$, if the created electron's spin align with the localized spin in $t_{2g}$. The total spin of this particular four-particles state is 2. On the other hand, the second line of Eq. \eqref{eqn:H0d-} describes a mixture of different 4-particle states with total spin 1. This implies that $\mathsf{d}^\dagger_{\dot{\mathbf{r}}\dot{\alpha},-}|\tilde{\Psi}\rangle$ is not the eigenstate of $\mathcal{H}_0$ due to the Hund's interaction. However, to facilitate the $C_n$ calculation, we decompose $\mathsf{d}^{\dagger}_{ \dot{\mathbf{r}}\dot{\alpha}-}|\tilde{\Psi}\rangle$ into the four-particle eigenstates with total spin 1. To find these eigenstates (last row in Table \ref{tbl:eigenenergy}), we solve the eigenvalue $\lambda^m_{\dot{\alpha}}$ and the eigenvectors $v^m_{k,\dot{\alpha}}$ in the following equation
\begin{equation}\label{eqn:v-lambda}
\Big[-\sum_{ \dot{\alpha}'\neq \dot{\alpha}}J^{ \dot{\alpha} \dot{\alpha}'}_{\dot{\mathbf{r}}}\hat{s}_{\dot{\mathbf{r}}\dot{\alpha}}\cdot \hat{s}_{\dot{\mathbf{r}}_1\dot{\alpha}'}-\lambda^m_{\dot{\alpha}}\Big]\sum_{k=0}^3v^m_{k,\dot{\alpha}}\phi^{k}_{\dot{\mathbf{r}}\dot{\alpha}}|\tilde{\Psi}\rangle=0,
\end{equation}
where $\phi^{0\dagger}_{ \dot{\mathbf{r}} \dot{\alpha}}=\mathsf{d}^{\dagger}_{\dot{\mathbf{r}}\dot{\alpha}-}$ and $\phi^{k,\dagger}_{ \dot{\mathbf{r}} \dot{\alpha}}=\mathsf{d}^{\dagger}_{\dot{\mathbf{r}}\dot{\alpha}+}\mathsf{d}^{\dagger}_{\dot{\mathbf{r}}k-}\mathsf{d}_{\dot{\mathbf{r}}k+}$ ($k=1,2,3$) are the operators that create the four-particle states with total spin 1. The index $m=0,\dots,3$
labels these excited eigenstates. Equation \eqref{eqn:v-lambda} is constructed by using the fact that the Hund's interaction preserves the total spin in the four-particle states. We shall not pursue this discussion any further, since these states are very likely to be the high-energy states for $\mathrm{CrX}_3$. We leave the details of these excitations in Appendix \ref{app:E_1p}. Here, we summarize the spectrum of these low-energy quasiparticles excitations in Table. \ref{tbl:eigenenergy}. With these excitation energies, we can proceed to evaluate the eigenvalue of $\mathrm{e}^{\tau \mathcal{H}_0}$ which are necessary for calculating $C_n$.

\subsection{Energy corrections due to hopping}
Once the hopping between lattices is allowed, this leads to the energy correction of the $\mathcal{H}_0$ ground state. This correction is obtained by evaluating $\mu_n$ in the correlation function $C_n$ [Eq. \eqref{eqn:connect}]. The evaluation of $\mu_n$ in Eq. \eqref{eqn:Gc} contains many hopping terms, where each of these terms corresponds to an evaluation of superexchange with a distinct hopping process. However, many of these hopping terms are zero, except for those having a hopping process with the ground state as its final state. This hopping process traces a \textit{closed} path on the lattice. Although we can discard many open-path processes immediately, $\mu_n$ still contains many hopping terms, particularly the processes that create several \textit{disconnected} closed-paths on the lattice. Although these processes have nonzero contributions, they cancel out exactly in Eq. \eqref{eqn:Gc}. This cancellation is guaranteed by linked-cluster theorem\cite{Metzner:PRB43(1991)} (see Appendix \ref{app:link-cluster}). Therefore, evaluating $C_n$ is reduced to calculating the hopping terms with the \textit{connected} closed-path. Despite this simplification, the calculation is still laborious, and we leave those details in Appendix \ref{app:states}. In the following, we only present the result for the spin-dependent part of $C_n$ up to $n=6$ since these are the nontrivial magnetic exchange energies at low-order corrections.

Keeping only the leading term (linear in $\beta$) in the low-temperature limit, we arrive at the first important result
\begin{equation}\label{eqn:C4text}
\frac{1}{2!}C_2+\frac{1}{4!}C_4=\beta \sum_{\dot{\mathbf{r}}_1\dot{\mathbf{r}}_2}\mathcal{J}_{\dot{\mathbf{r}}_1\dot{\mathbf{r}}_2}\mathbf{s}_{ \dot{\mathbf{r}}_1}\cdot\mathbf{s}_{ \dot{\mathbf{r}}_2}\\
\end{equation} with the Heisenberg exchange coupling
\begin{align}\label{eqn:J}
&\mathcal{J}_{\dot{\mathbf{r}}_1\dot{\mathbf{r}}_2}\!=\tfrac{\dot{\sigma}_1}{2}\Gamma^{m_1}_{ \dot{\alpha}_1\dot{\sigma}_1}\Big\{\frac{P_{\dot{\alpha}_2}u_{\dot{1}\dot{2}}u_{\dot{2}\dot{1}}}{\Omega^{m_1}_{ \dot{1}\dot{\sigma}_1}-\bar{\omega}_{ \dot{2}}}+
\frac{1}{\mathcal{E}^{m_1}_{ \dot{1} 1}}\Big[\frac{P_{ \dot{\alpha}_2}  t_{\dot{ 1} 1} t_{ 1 \dot{2}}t_{\dot{1} 2 } t_{ 2 \dot{2}}}{(\Omega^{m_1}_{ \dot{1}\dot{\sigma}_1}-\bar{\omega}_{ \dot{2}})\mathcal{E}^{m_1}_{ \dot{1} 2}}
\notag\\
&
\!-\!
\frac{\frac{\dot{\sigma}_2}{2}\Gamma^{m_2}_{ \dot{\alpha}_2\dot{\sigma}_2}  t_{\dot{1}1 } t_{\dot{2}2 }}{1\!+\!\Theta^{\eta}_{12}/(\mathcal{E}^{m_1}_{ \dot{1}1}\!+\!\mathcal{E}^{m_2}_{ \dot{2}2})}
\Big(
\frac{
  t_{ 2 \dot{1}}  t_{ 1 \dot{2}}}{
\mathcal{E}^{m_1}_{ \dot{1} 2}\mathcal{E}^{m_2}_{ \dot{2} 1}}
\!-\!\eta \frac{ t_{ 2 \dot{2} }  t_{ 1 \dot{1}} }{\mathcal{E}^{m_1}_{ \dot{1}  1}\mathcal{E}^{m_2}_{ \dot{2} 2}}
\Big)\Big]\Big\}.
\end{align}
$P_{\dot{\alpha}}=1$ for $\dot{\alpha}=1,2,3$ and zero otherwise. Here, we simplify the expression by letting the lattice and orbital indices as $n\equiv \mathbf{r}_n,\alpha_n$ and $\dot{n}\equiv \dot{\mathbf{r}}_n,\dot{\alpha}_n$. Furthermore, except $\dot{\mathbf{r}}_{1,2}$ (in the left-hand-side), summing all over the other indices in the right-hand-side in Eq. \eqref{eqn:J} is implicitly assumed. We will use this notation through out the text. In Eq. \eqref{eqn:J}, the denominator 
\begin{equation}
\mathcal{E}^{m}_{\dot{n}n}=\Omega^m_{\dot{\mathbf{r}}_n\dot{\alpha}_n\dot{\sigma}_n}-\bar{\nu}_{\mathbf{r}_n \alpha_n},
\end{equation}
is the energy for creating a $d$-electron and $p$-hole pair which is the elementary excitation energy due to the Cr-X hopping (Fig. \ref{fig:hopping}b). In the above, we have rewritten the $d$-orbital one-electron energies as
\begin{equation}\label{eqn:Omega}
\Omega^m_{\dot{1}\dot{\sigma}}=\begin{cases}
\delta_{\dot{\sigma},-} \omega_{\dot{1}},& \dot{\alpha}=1,2,3\\
\delta_{\dot{\sigma},+}\omega_{\dot{1}}+\delta_{\dot{\sigma},-}\varpi^m_{\dot{1}},& \dot{\alpha}=4,5
\end{cases}.
\end{equation}
We remind that the quaisparticle energies $\omega_{\dot{n}}$ and $\bar{\nu}_{n}$ are given in Table \ref{tbl:eigenenergy}.
The prefactor is
\begin{equation}\label{eqn:Gamma}
\Gamma^m_{ \dot{\alpha}\dot{\sigma}}=
\begin{cases}
\delta_{\dot{\sigma},-}\delta_{m,0},& \dot{\alpha}=1,2,3\\
\delta_{\dot{\sigma},+}\delta_{m,0}+\delta_{\dot{\sigma},-}|A^m_{\dot{\alpha}}|^2,& \dot{\alpha}=4,5
\end{cases},
\end{equation}
where $A^m_{\dot{\alpha}}$ is obtained by solving $\sum_{m=0}^3A^m_{\dot{\alpha}}v^m_{k,\dot{\alpha}}=\delta_{0,k}$ for $k=0,\dots3$.

In Eq. \eqref{eqn:J}, we may identify the first, second, and third terms as three distinct superexchange mechanisms according to the number of $p$ holes in the intermediate states. SE0 is the lowest-order superexchange which does not involve any $p$ orbitals. SE1 [Fig. \ref{fig:hopping}(c) ] and SE2 [Fig. \ref{fig:hopping}(d) - (f)] are the superexchange that involve one $p$ hole and two $p$ holes in the intermediate states. In SE0 and SE1, the resulting exchange coupling is mostly determined by the structure of $d$ orbitals instead of $p$ orbitals. The SE0 and SE1 processes may realize the FM exchange mediated by  $t_{2g}$ - $e_{g}$ hopping or AFM exchange mediated by $t_{2g}$ - $t_{2g}$ hopping. To determine which exchange coupling is energetically favorable, it only depends on two properties in the $d$-orbital: the $t_{2g}$ - $e_{g}$ splitting and the onsite interacting strength.

In the SE2 process, if the two intermediate $p$ holes locate at different lattices [Fig.\ref{fig:hopping}(d)], this process is similar to SE0 and SE1 where the $d$-orbital determines the sign of the exchange coupling. However, once the intermediate hole-pair is created locally on the same X ion, the correlated $p$-hole pair may mediate AFM exchange through spin-singlet pair [Fig.\ref{fig:hopping}(e)] or FM exchange through spin-triplet pair [Fig.\ref{fig:hopping}(f)]. To identify the most favorable exchange process in SE2, the geometrical effects in orbital overlapping play a major role. In short, we remark that the material's FM or AFM exchange is the result of interplay between SE0, SE1, and SE2 processes. The resulting exchange coupling is sensitive to the materials' crystal-field splitting, orbital overlapping, and onsite interactions.

Similarly, we proceed to calculate the correlation functions due to the spin-orbit coupling effects. The lowest order correction due to spin-orbit coupling is  
\begin{align}
\frac{1}{5!}C_5&=\beta\sum_{\dot{\mathbf{r}}_1\dot{\mathbf{r}}_2}\bm{\mathcal{D}}_{\dot{\mathbf{r}}_1\dot{\mathbf{r}}_2}\cdot\mathbf{s}_{ \dot{\mathbf{r}}_1}\times\mathbf{s}_{ \dot{\mathbf{r}}_2},\label{eqn:C5text}
\end{align}
where the Dzyaloshinskii–Moriya (DM) interactions 
\begin{align}
&\mathcal{D}^j_{\dot{\mathbf{r}}_1\dot{\mathbf{r}}_2}=
i\Lambda_{\mathbf{r}_2j}^{\alpha_1 \alpha_2}\delta_{\mathbf{r}_1\mathbf{r}_2}\frac{
\dot{\sigma}_1\Gamma^{m_1}_{\dot{\alpha}_1\dot{\sigma}_1}}{\mathcal{E}^{m_1}_{\dot{1}2}}\Big\{\frac{P_{ \dot{\alpha}_2}t_{ 3\dot{1}} t_{\dot{2} 3  }
    t_{\dot{1}  1} t_{2\dot{2}}}{\mathcal{E}^{m_1}_{\dot{1}1}\mathcal{E}^{m_1}_{\dot{1}3}(\Omega^{m_1}_{ \dot{1}\dot{\sigma}_1}-\bar{\omega}_{\dot{2}})}\notag\\
&
-\frac{\frac{1}{2}\dot{\sigma}_2\Gamma^{m_2}_{\dot{\alpha}_2\dot{\sigma}_2}t_{ 3\dot{2}}t_{ 2\dot{1}}/\mathcal{E}^{m_1}_{\dot{2}3}}{1+\Theta^{-\eta}_{23}/(\mathcal{E}^{m_1}_{\dot{1}2}+\mathcal{E}^{m_2}_{\dot{2}3})}
\Big[
\frac{t_{\dot{1}  3}  t_{\dot{2} 1}/\mathcal{E}^{m_2}_{\dot{2}1}-\eta t_{\dot{1}  1} t_{\dot{2} 3}/\mathcal{E}^{m_1}_{\dot{1}1}}{\mathcal{E}^{m_1}_{\dot{1}1}+\mathcal{E}^{m_2}_{\dot{2}3}+\Theta^{\eta}_{13}}
\notag\\
&+\frac{ t_{\dot{2}  1} t_{ \dot{1}3}}{\mathcal{E}^{m_2}_{\dot{2}1}\mathcal{E}^{m_2}_{\dot{2}2}}-\eta\frac{t_{\dot{1}  1} t_{ \dot{2}3}}{\mathcal{E}^{m_1}_{\dot{1}1}\mathcal{E}^{m_1}_{\dot{1}2}}\Big]\Big\}\label{eqn:D}
\end{align}
with $\Lambda_{\mathbf{r},j}^{\alpha_1 \alpha_2}=i\sum_{\alpha}\lambda^\alpha\mathcal{R}^{\mathbf{r}}_{ j \alpha}\varepsilon_{\alpha\alpha_2 \alpha_1}$.
$\mathcal{R}^{\mathbf{r}}_{j \alpha}$ is the rotational matrix that transform the local $\bar{x}\bar{y}\bar{z}$-system at $\mathbf{r}$ to the global $xyz$-system (see Fig.\ref{fig:1L-CrX3} and Appendix \ref{app:states}). We note that $j=x,y,z$ is to label the axes in $xyz$-system instead of labeling the $p$-orbital. This correction gives rise to the antisymmetric exchange which favors the formation of magnetic chiral order. The next order correction due to spin-orbit coupling is
\begin{equation}
\frac{1}{6!}C_6=\beta\sum_{\dot{\mathbf{r}}_1\dot{\mathbf{r}}_2}\sum_{ij}\mathcal{K}_{ \dot{\mathbf{r}}_1\dot{\mathbf{r}}_2}^{i j}s_{ \dot{\mathbf{r}}_1}^{i}s_{ \dot{\mathbf{r}}_2}^{j},\label{eqn:C6text}
\end{equation}
where $\mathcal{K}^{ij}_{\dot{\mathbf{r}}_1\dot{\mathbf{r}}_2}\propto(\lambda^\alpha)^2$ is a symmetric tensor in $i$ and $j$ indices. This term is the symmetric exchange interaction that determines the magnetic anisotropy exchange (MAE) and the Kitaev interactions\cite{Xu:npjCompMat4(2018),Xu:PRL124(2020),Lee:124(2020)}. This small correction is important to stabilize the magnetic order in 2D system\cite{Lado:2DMat4(2017)}. The calculation of this higher-order exchange process is tedious. Therefore, we omit the derivation of $\mathcal{K}$ and outline the steps in Appendix \ref{app:states} for the interested reader.

Next, we discuss the inter-layer superexchange which is one of the most interesting aspects in vdW 2D magnets. Using the similar approach, it is straightforward to generalize the calculation to a bilayer system with arbitrary hopping processes. To do this, we let the inter-layer hopping Hamiltonian as
\begin{equation}
\mathcal{H}_{\perp}=\sum_{\ell\neq\ell'}T^{\ell,\ell'}_{\mathbf{r}\alpha,\mathbf{r}'\alpha'}p^{\ell\dagger}_{\mathbf{r}\alpha \sigma}p^{\ell'}_{\mathbf{r}'\alpha' \sigma},
\end{equation}
where $\ell=1,2$ is the layer index. The lowest order corrections for the exchange energy due to this hopping is $n=6$,
\begin{equation}\label{eqn:C6perp}
\frac{1}{6!}C^\perp_6=\beta \sum_{\ell_1\neq\ell_2}\sum_{ \dot{\mathbf{r}}_1\dot{\mathbf{r}}_2}\tilde{\mathcal{J}}^{ \ell_1\ell_2}_{ \dot{\mathbf{r}}_1\dot{\mathbf{r}}_2}\mathbf{s}^{\ell_1}_{\dot{\mathbf{r}}_1}\cdot\mathbf{s}^{\ell_2}_{\dot{\mathbf{r}}_2}
\end{equation}
with the inter-layer exchange coupling
\begin{align}
\tilde{\mathcal{J}}^{ \ell_1\ell_2}_{ \dot{\mathbf{r}}_1\dot{\mathbf{r}}_2}\!&=\!
\frac{\frac{\dot{\sigma}_1}{2}\Gamma^{m_1}_{\dot{\alpha}_1\dot{\sigma}_1}t_{\dot{1}  1}t_{  4\dot{1}}}{\mathcal{E}^{m_1}_{\dot{1}1}\mathcal{E}^{m_1}_{\dot{1}4}}\Big\{\frac{P_{\dot{\alpha}_2}T^{\ell_2\ell_1}_{ 34 } t_{ \dot{ 2}3 }
T^{\ell_1\ell_2}_{ 1 2}  t_{ 2 \dot{ 2}}}{\mathcal{E}^{m_1}_{\dot{1}2}\mathcal{E}^{m_1}_{\dot{1}3}(\Omega^{m_1}_{\dot{1}\dot{\sigma}_1}-\bar{\omega}_{\dot{2}})}\!-\!\tfrac{\dot{\sigma}_2}{2}\Gamma^{m_2}_{\dot{\alpha}_2\dot{\sigma}_2}\notag\\
&\frac{t_{\dot{2} 2} T^{\ell_1\ell_2}_{ 13}-\eta t_{\dot{2} 3} T^{\ell_1\ell_2}_{ 12}}{\mathcal{E}^{m_1}_{\dot{1}3}\!\!+\!\mathcal{E}^{m_2}_{\dot{2}2}\!\!+\!\Theta^\eta_{23}}
\Big[
 \frac{t_{ 2\dot{2}} T^{\ell_2\ell_1}_{ 34}}{(\mathcal{E}^{m_1}_{\dot{1}3})^2}+\frac{t_{ 2\dot{2}}  T^{\ell_2\ell_1}_{ 34}}{(\mathcal{E}^{m_2}_{\dot{2}2})^2}
\Big]\Big\}.\label{eqn:J_inter}
\end{align}
The exchange mechanism in the first term is similar to SE1. It does not involve any intermediate hole-pairs in the $p$-orbitals. As we will see later, this SE1 process only gives rise to inter-layer FM exchange in $\mathrm{CrX}_3$ regardless of the stacking order. On the contrary, the second term corresponds to the SE2 processes involving the intermediate correlated hole-pairs. This SE2 process may realize the inter-layer AFM coupling by the singlet-pair-mediated superexchange.

In summary, we have constructed a microscopic model for the magnetic ground state of $\mathrm{CrX}_3$ by projecting the many-body state onto the Mott state. We go beyond the lowest order correction ($n=2$) to the ground state and take into account the higher-order corrections (up to $n=6$) due to the virtual processes in Cr-X-Cr hopping, Cr-X-X-Cr hopping, and SOC. This yields various types of exchange coupling constants which can be explicitly expressed in terms of the TB constants and onsite interactions. Collecting all the results in Eqs. \eqref{eqn:C4text}, \eqref{eqn:C5text}, and \eqref{eqn:C6perp}, we finally arrive at the effective spin model in Eq. \eqref{eqn:connect} for the many-body system. Our next task is to estimate these microscopic model parameters in Eqs. \eqref{eqn:J}, \eqref{eqn:D} and \eqref{eqn:J_inter} which can be extracted from the quasiparticle spectrum in the first-principles studies. This will be the topic in the next section.

\section{Application to Monolayer $\mathrm{CrI}_3$}\label{sec:DFT}

In this section, we apply our superexchange model to $\mathrm{CrI}_3$. To investigate its magnetic ground state, we use the \textit{ab initio} method to estimate the model parameters in Eqs. \eqref{eqn:J}, \eqref{eqn:D} and \eqref{eqn:J_inter}. 

\subsection{DFT calculation}
The electronic structure of $\mathrm{CrI}_3$ is calculated by using QUANTUM ESPRESSO\cite{Giannozzi:JPhysCondMat29(2017)}. In the calculation, we use the Perdew-Burke-Ernzerhof (PBE) pseudopotentials from the standard solid-state pseudopotential efficiency library\cite{Lejaeghere:Science351(2016),Prandini:njpCompMat4(2018)}. The in-plane lattice parameter is adopted from the experimental value\cite{McGuire:ACScm27(2015)} 6.867 \AA. To eliminate interaction between supercells, we introduce more than 20 \AA\ vacuum between the supercell images in the out-of-plane direction. We perform the calculation with a fixed unit-cell volume and the lattice structure is optimized by the relax calculation until the atomic residual force is less than $3\times10^{-4}$ eV/\AA. The numerical integration over the Brillouin zone is done by sampled over $8\times8\times1$ $\Gamma$-centered Monkhorst-Pack grid. The energy cutoff for wavefunction and density is 50 Ry and 450 Ry. 
In this simulation, we find the nearest-neighbor exchange interaction: $2.49$ meV \footnote{The exchange interaction $\mathcal{J}$ is obtained by using $E_{\text{AFM}}-E_{\text{FM}}=\frac{2}{N_{\text{c}}}\sum_{\langle ij\rangle}\mathcal{J}|S|^2$ where $N_{\text{c}}$ is the number of unit-cell and $S=3/2$ is the spin. The total energy per unit cell of antiferromagnetic and ferromagnetic ground states from DFT calculation are denoted as $E_{\text{AFM}}$ and $E_{\text{FM}}$ (see Refs. \cite{Zhang:JMatChemC3(2015),Lado:2DMat4(2017)}).}
and magnetization: $2.95\mu_B$/Cr which are consistent with the previous studies\cite{Zhang:JMatChemC3(2015),Lado:2DMat4(2017),Besbes:PRB99(2019),Torelli:2DMat6(2019),Wu:PhysChemChemPhys21(2019),Kashin:2DMat7(2020),Stavropoulos:PRR3(2021)}. To estimate the TB constants, we construct the maximally localized Wannier functions\cite{Marzari:RMP84(2012)} by using WANNIER90 \cite{Pizzi:JPhysConfMat32(2020)}. In the calculation, the $p$ and $d$ Wannier orbitals are projected by using the $xyz$-coordinate system (see Fig. \ref{fig:1L-CrX3}). We summarize the TB constants obtained from WANNIER90 in Appendix  \ref{app:TB}.
\begin{table}
\caption{\label{tbl:DFTpara}The quasiparticle excitations extracted from DFT calculation ($E_F=-3.3806$ eV is the Fermi energy). Unless stated, the unit is in eV. In this table, the spin of the $d$-hole and electron is always parallel to the localized spins in $t_{2g}$ ($\dot{\sigma}=+1$). The high-energy antiparallel spin $d$-electron states are ignored. The scissor corrected results are given in the parenthesis.}
\begin{ruledtabular}
\begin{tabular}{lccc}
Excitations: & States: & &\\
\hline
$p$-hole : & $\alpha=p_z,$ & $p_y,$ & $p_x$
\\
$\bar{\nu}_{\mathbf{r}\alpha}$ & $-4.289$, & $-4.897$, & $-5.044$
\\
\hline
$d$-hole ($a_{1g}$, $e^{\pi}_g$): & $\dot{\alpha}=1,$ & 2, & 3
\\
$\bar{\omega}_{\dot{\mathbf{r}}\dot{\alpha}}$&$-4.156$, &$-4.126$,&$-4.125$
\\
$d$-electron ($e^{\sigma}_g$): & $\dot{\alpha}=4,$ & 5, & \\
$\omega_{\dot{\mathbf{r}}\dot{\alpha}}$ &$-3.183$, &$ -3.182$& \\
(scissor corrected) &$(-1.413)$, &$ (-1.412)$& \\
\hline
Exchange coupling: $\mathcal{J}$ & $J_p=0.5$ & $J_p=0$&\\
from Eq.\eqref{eqn:J}  &$U_p=1.2$&$U_p=0$&\\
&$29.1$meV&$17.1$meV&\\
(scissor corrected) &($2.53$meV), &($1.73$meV) &
\end{tabular}
\end{ruledtabular}
\end{table}

\subsection{quasiparticle spectrum}
From the DFT quasiparticle dispersion, we can extract the excitation energies of holes and electrons in the $p$ and $d$ orbitals. The result is summarized in Table \ref{tbl:DFTpara}. We note that these quasiparticle energies in DFT calculation have already accounted for the interactions. This effect leads to the evident splitting in the spin-up and spin-down onsite energies of the Wannier TB Hamiltonian. Instead of identifying them as the non-interacting onsite energy $\epsilon^\alpha_{\mathbf{R}}$ (spin-independent), these computed band energies should correspond to the quasiparticle energies: $\bar{\omega}_{\dot{\mathbf{r}}\dot{\alpha}}$, $\omega_{\dot{\mathbf{r}}\dot{\alpha}}$, and $\bar{\nu}_{\mathbf{r}\alpha}$ in Eqs. \eqref{eqn:J}, \eqref{eqn:D} and \eqref{eqn:J_inter}. One may  extract the onsite energy and coupling constants from the spectrum with further analysis\cite{Yekta:PRMat5(2021),Soriano:njpCompMat7(2021)}, but this is not the main point of our study. One may also note in Table \ref{tbl:DFTpara} that. The table does not include all the possible intermediate excitations in Eqs. \eqref{eqn:J}, \eqref{eqn:D} and \eqref{eqn:J_inter}. In particular, we exclude the excitations in the $d$ orbital with spin antiparallel ($\dot{\sigma}=-1$) to the localized spins in the $t_{2g}$, since our DFT-computed band dispersion shows that these excitations are at least 2eV larger than parallel spin excitations. This may be due to the strong onsite Hubbard-$U$ interaction with a relatively small $t_{2g}-e_g$ splitting. Therefore, we only keep the excitations in creating an electron with $\dot{\sigma}=+1$ and $\dot{\alpha}=4,5$ on Cr since the other high-energy excitations are suppressed by a large denominator in Eq. \eqref{eqn:J}, \eqref{eqn:D} and \eqref{eqn:J_inter}.

Identifying the Kohn-Sham spectrum as the quasiparticle band structure may lead to underestimating the band gap. In our DFT calculation, the band gap in the quasiparticle dispersion is approximately $1$ eV ($t_{2g}$-$e_g$ splitting). This value is well below to that in the $GW$ calculation \cite{Wu:NatComm10(2019),Acharya:PRB104(2021)} ($\approx3$ eV). To remedy this problem, we may employ the `scissor' correction\cite{Fiorentini:PRB51(1995),Johnson:PRB58(1998),Bernstein:PRB66(2002),Parashari:PhysicaB403(2008),Thilagam:JPhysCondMat23(2010),Babu:JSSChem184(2011),Magorrian:PRB94(2016)} to shift all the conduction bands rigidly up above the valence band by an additional $1.77$ eV as suggested by the LDSA and $GW$ calculation in Ref.\cite{Wu:NatComm10(2019)}. This band gap correction is crucial to obtaining the exchange couplings with the correct order of magnitude. To see this, we estimate the exchange coupling by using \eqref{eqn:J} and the quasiparticle energy in Table \ref{tbl:DFTpara}. Without the scissor correction, the exchange constant $\mathcal{J}$ is overestimated by an order of magnitude (Table \ref{tbl:DFTpara}).

\subsection{Correlation effects in $p$ orbital}
\begin{figure}
\centering
\includegraphics[width=3.3in]{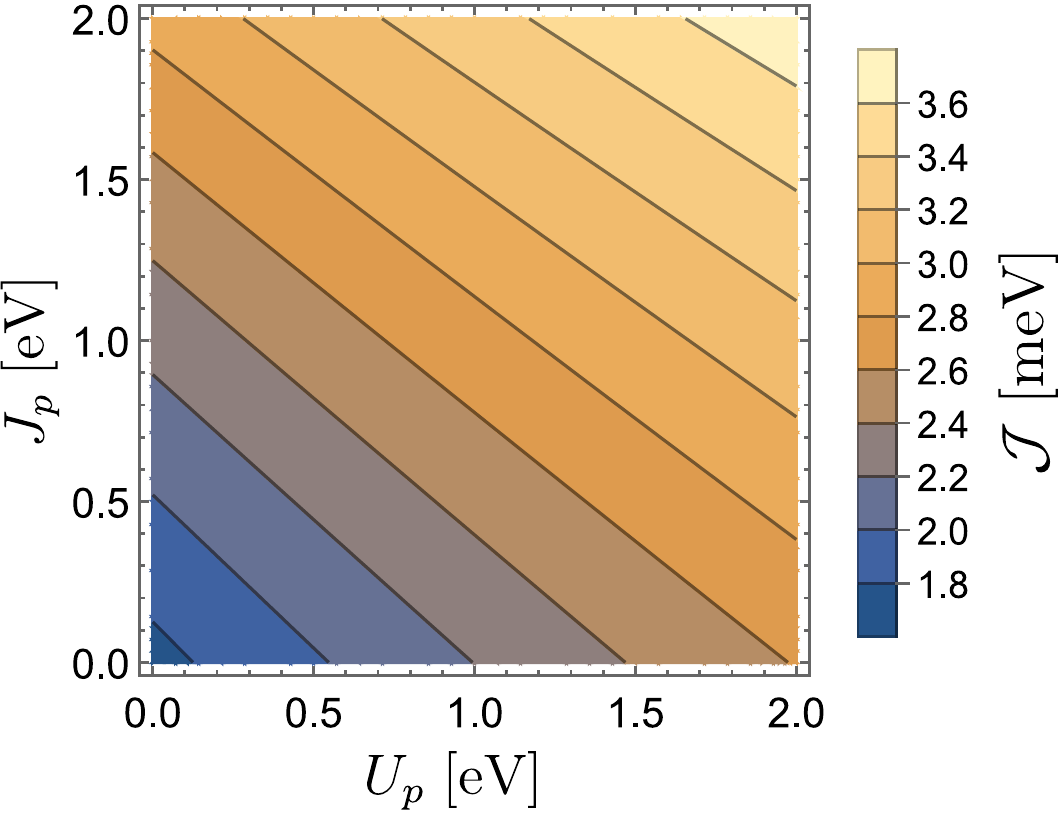}
\caption{The $U_p$ - $J_p$ parameter space for the intra-layer exchange coupling ($\mathcal{J}$). As we can see, in the weak correlation regime (left bottom corner), the exchange coupling is signifcantly lower than the DFT calculated value ($\sim2.5$meV).}\label{fig:U-J}
\end{figure}

A final piece of important information is the energy splitting between the spin-singlet and spin-triplet states in the $p$-orbital. This energy splitting is determined by the two-particle spectrum which we cannot obtain from the Kohn-Sham quasiparticle dispersion. Therefore, the $p$-orbital Hund's interaction $J^{\alpha\alpha'}_{\mathbf{r}}$ and Hubbard interaction $U^{\alpha\alpha'}_{\mathbf{r}}$ remain the free parameters in our model. To see how this singlet-triplet splitting influences the exchange coupling $\mathcal{J}$, we let $U^{\alpha\alpha'}_{\mathbf{r}}=U_p$ and $J^{\alpha\alpha'}_{\mathbf{r}}=J_p$ plot in Fig. \ref{fig:U-J}. We found that the exchange coupling from our model agrees well with the DFT and experimental value within the reasonable range of $U_p$ and $J_p$. For instance, using $J_p=0.5$eV and $U_p=1.2$eV, we can obtain the exchange constant at $2.53$meV.
Furthermore, we also check the case with smaller interorbital interaction: $U^{\alpha\alpha}_{\mathbf{r}}=U_p>U^{\alpha\neq\alpha'}_{\mathbf{r}}$ and $J^{\alpha\alpha'}_{\mathbf{r}}=J_p$.  This enhances the singlet-triplet splitting leading to stronger exchange interaction.

\subsection{Intra-layer exchange}
\begin{figure}
\centering
\includegraphics[width=3.3in]{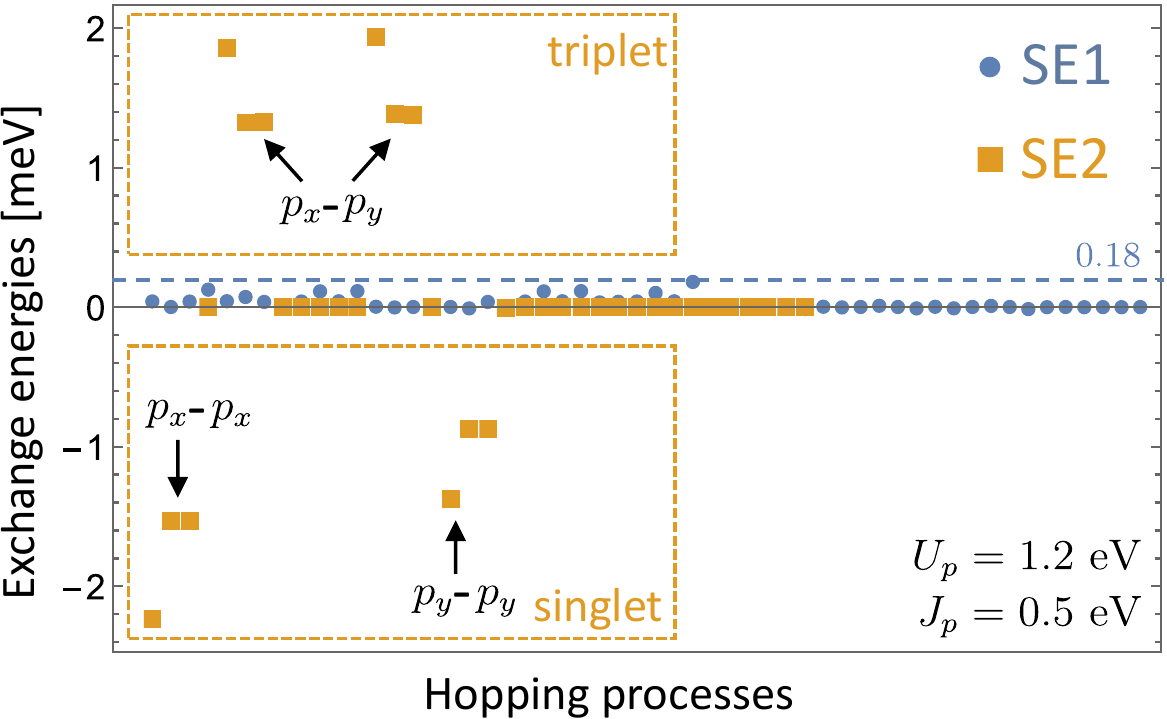}
\caption{The exchange couplings that arise from each distinct hopping processes. The exchange energy contributions for each hopping processes in SE1 is not larger than $0.18$meV (blue dashed line). The largest SE1 contribution (0.18 meV) comes from the $\dot{\alpha}=5$ to $\dot{\alpha}=3$ hopping via $p_y$ as the intermediate state. In general, SE2 has larger contribution to the exchange coupling than SE1. But, the exchange coupling mediated by the singlet-pairs in $p_x$ and $p_y$ orbital (lower dashed box) are canceled out with the triplet-pairs in $p_x$-$p_y$ orbital (upper dashed box).}\label{fig:channels}
\end{figure}

The FM superexchange in $\mathrm{CrI}_3$ is commonly attributed to its approximately $90^{\circ}$ Cr-I-Cr bond angle. In this geometry, the hopping amplitude in SE1 may be suppressed due to the lack of orbital overlapping, such that, the higher energy SE2 becomes the dominant process. Typically, the SE2 process in this geometry favors the FM superexchange because of the Hund interaction in the intermediate spin-triplet hole-pairs. This is the celebrated Goodenough-Kanamori (GK) rule\cite{Goodenough:JPhysChem6(1958),Kanamori:JPhysChem10(1959)}. However, we argue that the intra-layer exchange interaction is a result of complicated competition between different intermediate states in $\mathrm{CrI}_3$. The full picture of the emergence of FM exchange in $\mathrm{CrI}_3$ is more subtle than the simple picture from GK's $90^{\circ}$ rule.

First, we note in our model that the direct Cr-Cr hopping contributes $0.5$ meV out of 2.53 meV to the FM exchange. This contribution is not negligibly small. Second, we observe that the exchange constant without spin singlet-triplet splitting in the $p$ orbital, $U_p=J_p=0$, the predicted exchange coupling (1.73 meV) is well below the DFT computed-value (2.49 meV). This implies that SE1 and SE2 have comparable contributions to the FM exchange, even though the individual SE1 process is weak (Fig. \ref{fig:channels}). Therefore, the intra-layer FM exchange is a complex result coming from all equally important contributions in SE0, SE1, and SE2 instead of predominantly from the SE2 as suggested by GK's $90^{\circ}$ rule. On the contrary, SE0 and SE1 together are the dominating FM exchange processes through the $t_{2g}$-$e_g$ hopping\cite{Ghosh:arXiv(2022)}.

Unexpected by GK's picture, we find that the AFM exchange coupling mediated by the (intra-orbital) spin-singlet hole-pair is as large as the FM exchange in the SE2 (see Fig. \ref{fig:channels}). Nevertheless, this result is consistent with the non-negligible SE1 contribution as discussed in the previous paragraph. This unprecedented result may attribute to the anisotropic geometry of the 2D materials such that the $p$ orbitals do not orient in the sense as described in a 3D bulk environment. Therefore, the singlet channel competes with the triplet channel leading to weak FM exchange (Fig. \ref{fig:channels}). This particular scenario may be unique to 2D materials. Therefore, the application of the GK rule to such a case may not yield the full picture of the underlying exchange mechanism.

\subsection{DM interaction}
The DM interaction may have important effects on the magnon physics\cite{Chen:PRX8(2018),Chen:PRX11(2021),Mook:PRX11(2021),Cai:PRB104(2021),Ghader:NJP23(2021)} in $\mathrm{CrX}_3$. In a perfect $\mathrm{CrX}_3$ crystal, the DM coupling vector $\bm{\mathcal{D}}$ is nearly zero due to inversion symmetry. The antisymmetric superexchange mediated by the upper X-layer cancel out with the lower X-layer\cite{Keffer:PR126(1962),Lado:2DMat4(2017)}(Fig. \ref{fig:1L-CrX3}). %
Nevertheless, to further verify our model, we also calculate the DM interacting strength. Using the spin-orbit coupling $\lambda^\alpha=0.6$ eV \cite{Kramida:NISTData(1999),Lado:2DMat4(2017)}, we obtain the DM coupling vector that mediated by the I ion at $h_1$: $\bm{\mathcal{D}}\approx[0.164,-0.136,0.242]$meV, and $h_2$: $\bm{\mathcal{D}}\approx-[0.163,-0.161,0.223]$meV (see Fig. \ref{fig:1L-CrX3}a). As expected, we find that these vectors canceled out due to the crystal symmetry. Interestingly, we find that the DM interaction is mostly mediated by the SE2 processes which are about four times larger than the SE1 contribution. This implies that the correlation effects of the intermediate $p$-hole pairs play an important role in this interaction. 

The DM interaction may be enhanced by applying electric fields\cite{Liu:PRB97(2018),Liu:AIP8(2018),JaeschkeUbiergo:PRB103(2021)}, distorting the lattice \cite{Ghosh:PhysicaB570(2019)}, synthesizing the Janus structure \cite{Albaridy:JPhysCondMat32(2020)}, and fabricating heterobilayer\cite{Pang:JPhysChemC125(2021)}. Furthermore, there are several studies suggested that the next-nearest anti-symmetry exchange may give nonzero DM interaction\cite{Kvashnin:PRB102(2020),JaeschkeUbiergo:PRB103(2021)}. In this higher-order exchange coupling, the effects of direct hopping between $p$ orbitals may no longer be negligible. This topic deserves further studies in the future.

\subsection{Interlayer exchange}
To investigate the inter-layer exchange, we may use the technique in Refs. \cite{Guinea:PRB99(2019), Koshino:PRX8(2018), Sboychakov:RPB92(2015), Moon:PRB87(2013), Laissardiere:NanoLett10(2010), Tang:PRB53(1996),Fang:PRB92(2015)} to model the inter-layer hopping strength as
\begin{equation}
T^{\ell,\ell'}_{\mathbf{r}\alpha,\mathbf{r}'\alpha'}\!=\!\sum_{jj'}\mathcal{R}_{j \alpha}^{\mathbf{r},\dagger}\Big[\delta_{j j'}V_\pi^{\perp}\!+\!(V_\sigma^{\perp}\!-\!V_\pi^{\perp})\frac{x_jx_{j'}}{x^2}\Big]\mathcal{R}_{j' \alpha'}^{\mathbf{r}}
\end{equation}
where $\bm{x}=\mathbf{r}+\mathbf{d}_{\ell}-(\mathbf{r}'+\mathbf{d}_{\ell'})$ being displacement between two X ions, and
$\mathbf{d}_{1,2}=\pm \frac{1}{2}(0,0,d)$ with $d$ being the inter-layer distance. We note that the Slater-Koster (SK) overlap integral $V^{\perp}_\sigma$ and $V^{\perp}_\pi$ are also a $\bm{x}$-dependent functions\cite{Tang:PRB53(1996)}.

In contrast to intra-layer exchange, the inter-layer exchange coupling is a super-superexchange (SSE) of Cr-I-I-Cr hopping. The relevant hopping processes are much larger than Cr-I-Cr hopping. The GK rule does not apply to the inter-layer SSE. The detailed quantitative study of this hopping effects on the inter-layer superexchange deserves a separate publication, especially developing the quantitative model for the SK overlap integrals\cite{Guinea:PRB99(2019), Koshino:PRX8(2018), Sboychakov:RPB92(2015), Moon:PRB87(2013), Laissardiere:NanoLett10(2010), Tang:PRB53(1996)}. Therefore, we leave them for future studies.

Nevertheless, without a quantitative analysis, we can still draw useful conclusions from our theory for the inter-layer exchange. First, similar to the intra-layer exchange, inter-layer SE1 processes that involve $t_{2g}$-$e_{g}$ hopping can only yield FM exchange due to strong Hubbard-$U$  interaction with relatively small $t_{2g}$-$e_g$ splitting. Thus, the inter-layer AFM exchange must arise from the SE2 mediated by the spin singlet-hole pair in the $p$ orbitals. This SE2 process includes the $e_{g}$-$e_{g}$, $t_{2g}$-$e_g$, and $t_{2g}$-$t_{2g}$ hopping. However, our intra-layer analysis suggests that occupying $t_{2g}$ with an additional electron is highly unlikely due to the strong Hubbard-$U$ interactions. This leads to strong suppression in $t_{2g}$-$e_g$, and $t_{2g}$-$t_{2g}$ hopping processes. This result agrees with the finding in DFT studies\cite{Jang:PRM3(2019), Sivadas:ACSnano18(2018),Yu:APLett119(2021)}.

\section{Conclusion}\label{sec:conclusion}
In conclusion, we have built a microscopic model for the superexchange in monolayer and bilayer $\mathrm{CrX}_3$. In particular, our model accounts for the superexchange mediated by the correlated hole pair (SE2) in the X ion that has not been considered in the previous studies. With this model, we derive the higher-order magnetic exchange processes including intra-layer DM interaction (nearest-neighbor) in Eq. \eqref{eqn:D} and the general form of inter-layer exchange coupling presented in Eq. \eqref{eqn:J_inter}.

In our DFT study of monolayer $\mathrm{CrI}_3$, we found that the SE2 process not only contributes substantially to the intra-layer FM exchange, but also to DM interaction. Interestingly, we found that the superexchanges mediated by the singlet and triplet hole pairs are comparable, but they cancel out each other resulting in weak intra-layer FM exchange coupling. We argue that the GK's $90^{\circ}$ rule does not yield the complete picture for this particular 2D material due to the nontrivial competing effects between different comparable superexchange processes.

In the bilayer $\mathrm{CrX}_3$, we demonstrated that the SE1 process is not sufficient to explain the stacking-dependent magnetism since it always leads to inter-layer FM exchange. Hence, the inter-layer AFM exchange in $\mathrm{CrX}_3$ may be realized by the SE2 mediated by the spin singlet-hole pair in $p$ orbitals. Specifically, our analysis shows that $e_{g}$-$e_{g}$ hopping is the most relevant SE2 process mediating the interlayer AFM exchange. This qualitative result agrees with previous DFT studies\cite{Jang:PRM3(2019), Sivadas:ACSnano18(2018)}.

The developed theory shall facilitate the studies of other important higher-order magnetic exchange effects, such as the next-nearest neighbor DM interaction, inter-layer DM interactions, symmetric exchange, and four-spin interactions\cite{Soriano:NanoLett20(2020)}. Furthermore, this microscopic theory is also useful for the future studies in the excited states properties such as magneto-optics properties\cite{Sivadas:PRL117(2016),Zhang:NanoLett20(2019),Jin:NatComm11(2020),Tomarchio:SciRep11(2021),Lei:PRM5(2021),Kudlis:PRB104(2021),Dirnberger:NatNanotech17(2022)} in $\mathrm{CrX}_3$. The theoretical technique in this paper can also be applied to the other interesting 2D magnets\cite{Wang:2DMat3(2016),Tian:2DMat3(2016),Lee:NanoLett16(2016),Gong:Nature546(2017),Bonilla:NatNanotech13(2018),Kim:PRL120(2018),Wang:NatNanotech13(2018),Fei:NatMat17(2018)}.

\begin{acknowledgements}
The authors would like to thank Alexandru Georgescu and Marco Berritta for the simulating discussions in the first principle studies. K.W.S. is supported by UK EPSRC New Investigator Award under the Agreement No. EP/V00171X/1. V.F. acknowledge support from ERC Synergy Grant Hetero2D, EC Quantum Technologies Flagship Project No. 2D-SIPC, EPSRC Grant No. EP/N010345, and the Lloyd Register Foundation Nanotechnology grant. 
\end{acknowledgements}

\appendix

\section{Quasiparticle and the excitations}\label{app:E_1p}

To calculate the exchange energies, we need to evaluate the $\tau$-evolution ($\mathrm{e}^{\tau \mathcal{H}_0}$) of the intermediate excited states. To do this, we reduce the calculation to evaluating the excited eigenstates that are described in Sec \ref{sec:ExStates}. With these eigenstates at hand, the $\tau$-evolution of the intermediate states can be trivially obtained by replacing $\mathcal{H}_0$ in $\mathrm{e}^{\tau \mathcal{H}_0}$ by the corresponding eigenvalues. Generally, the intermediate excitations created by the hopping may not be the eigenstates of $\mathcal{H}_0$ such as the two-hole excitation
\begin{equation}
p_{\mathbf{r}_2 \alpha_2\sigma_2}p_{\mathbf{r}_1\alpha_1\sigma_1}|\tilde{\Psi}\rangle=\sum_{\eta=-1,1}\Phi^{\eta,\alpha_1\alpha_2}_{\mathbf{r}_1\mathbf{r}_2,\sigma_1\sigma_2}|\tilde{\Psi}\rangle,\label{eqn:2-h_decom}
\end{equation}
and the one-electron state
\begin{align}
\mathsf{d}^\dagger_{ \dot{\mathbf{r}} \dot{\alpha}\dot{\sigma}}&|\tilde{\Psi}\rangle= [\delta_{ \dot{\sigma},-}P_{ \dot{\alpha}}\!+\!\delta_{ \dot{\sigma},+}(1-P_{ \dot{\alpha}})]\mathsf{d}^{\dagger}_{ \dot{\mathbf{r}} \dot{\alpha} \dot{\sigma}}\notag\\
&+\delta_{ \dot{\sigma},-}(1-P_{ \dot{\alpha}})\sum_{m=0}^{3}A^{m}_{ \dot{\alpha} }\sum_{k=0}^3 v^{k}_{m,\dot{\alpha}}\phi^{k\dagger}_{ \dot{\mathbf{r}} \dot{\alpha}}|\tilde{\Psi}\rangle,\label{eqn:1-d_decom}
\end{align}
where $\sum_{m=0}^{3}A^{m}_{\dot{\alpha}}v^k_{m,\dot{\alpha}}=\delta_{k,0}$. Nevertheless, we can always expand them into the eigenstates in Table \ref{tbl:eigenenergy}.

Most of the eigenstates of $\mathcal{H}_0$ are straightforward to obtain, except the four-particle states in the second line of Eq. \eqref{eqn:1-d_decom}. As indicated in Eq. \eqref{eqn:H0d-}, $\mathsf{d}^{\dagger}_{ \dot{\mathbf{r}}\dot{\alpha}-}|\tilde{\Psi}\rangle$ with $\dot{\alpha}\in e_g$ is not the eigenstate of $\mathcal{H}_0$. Namely, the Hund interaction leads to the mixing of $\mathsf{d}^{\dagger}_{ \dot{\mathbf{r}}\dot{\alpha}-}|\tilde{\Psi}\rangle$ with the other different four-particle states having the same total spin, since the Hund interaction commutes with the total spin
\begin{equation}\label{eqn:[S,HJ]}
\Big[\sum_{\dot{\alpha} }\hat{s}_{\dot{\mathbf{r}} \dot{\alpha}},\sum_{ \dot{\alpha}'\neq \dot{\alpha}}J^{ \dot{\alpha} \dot{\alpha}'}_{\dot{\mathbf{r}}}\hat{s}_{\dot{\mathbf{r}}\dot{\alpha}}\cdot \hat{s}_{\dot{\mathbf{r}}_1\dot{\alpha}'}\Big]=0.
\end{equation}
This relation also implies that the full four-particle Hilbert space breaks into different independent eigenspace with the same total spin.

To facilitate the evaluation of $\tau$-evolution in $C_n$, we can expand Eq. \eqref{eqn:H0d-} into the four-particle eigenstates with total spin 1. To find these eigenstates, we only need to focus on the eigenspace with total spin 1. This particular eigenspace is completely spanned by the four states $\phi^{k}_{\dot{\alpha}}|\tilde{\Psi}\rangle$ with $\phi^{0\dagger}_{ \dot{\mathbf{r}} \dot{\alpha}}=\mathsf{d}^{\dagger}_{\dot{\mathbf{r}}\dot{\alpha}-}$ and $\phi^{k,\dagger}_{ \dot{\mathbf{r}} \dot{\alpha}}=\mathsf{d}^{\dagger}_{\dot{\mathbf{r}}\dot{\alpha}+}\mathsf{d}^{\dagger}_{\dot{\mathbf{r}}k-}\mathsf{d}_{\dot{\mathbf{r}}k+}$ ($k=1,2,3$). Thus, this gives the eigenvalue problem for $\mathcal{H}_0$ in Eq. \eqref{eqn:v-lambda}. After obtaining the eigenstates, we can expand $\mathsf{d}^{\dagger}_{ \dot{\mathbf{r}}\dot{\alpha}-}|\tilde{\Psi}\rangle$ into the four-particle eigenstates [second line in Eq. \eqref{eqn:1-d_decom}]. As we mentioned in the main text, these excited states may be ignored, since they are high-energy states and may not correspond to the one-particle excitation in the Kohn-Sham quasiparticles spectrum.

\section{$\mu_n$, $C_n$ and link-cluster theorem}\label{app:link-cluster}

We note in Eq. \eqref{eqn:Gc} that $\mu_n$ contains many hopping terms and each of them is described by a superexchange with a distinct hopping process. Many of these hopping terms do not contribute to $C_n$. To find the relevant nonzero hopping terms in $\mu_n$, we may write the product of $\mathcal{H}'$ as the sum of all position indices $\mathbf{r}$ as
\begin{align}\label{eqn:prod}
\prod^n_{j=1}\mathcal{H}'(\tau_j)=\sum_{\mathbf{r}_1\dots\mathbf{r}_n}\sum_{\mathbf{r}_1'\dots\mathbf{r}_n'}\mathcal{O}_{\mathbf{r}_n\mathbf{r}_n'}^{(\tau_n)}\dots\mathcal{O}_{\mathbf{r}_1\mathbf{r}_1'}^{(\tau_1)}
\end{align}
where
\begin{equation*}
\mathcal{O}_{\mathbf{r}\mathbf{r}'}^{(\tau)}=\begin{cases}
\displaystyle\sum_{\alpha \alpha'}\sum_{\sigma}p^\dagger_{\mathbf{r}\alpha \sigma}(\tau)t_{\mathbf{r}\alpha, \mathbf{r}'\alpha'}d_{\mathbf{r}'\alpha' \sigma}(\tau),\\
\displaystyle\sum_{\alpha \alpha'}\sum_{\sigma}d^\dagger_{\mathbf{r}\alpha \sigma}(\tau)t_{\mathbf{r}\alpha, \mathbf{r}'\alpha'}p_{\mathbf{r}'\alpha' \sigma}(\tau),\\
\displaystyle\sum_{\alpha \alpha'\alpha''}\sum_{\sigma \sigma'}\delta_{\mathbf{r} \mathbf{r}'}\Lambda^{\alpha''}_{\alpha \alpha'}\tau^{\alpha''}_{\sigma' \sigma'} p^\dagger_{\mathbf{r} \alpha \sigma}(\tau)p_{\mathbf{r}\alpha' \sigma'}(\tau').
\end{cases}
\end{equation*}
We say that $\mathcal{O}_{\mathbf{r}_1\mathbf{r}_1'}^{(\tau_1)}$ and $\mathcal{O}_{\mathbf{r}_2\mathbf{r}_2'}^{(\tau_2)}$ are \textit{connected} if at least one of the following is true: (1) $\mathbf{r}_1'=\mathbf{r}_2' \text{ or }\mathbf{r}_2$ (2) $\mathbf{r}_1=\mathbf{r}_2' \text{ or }\mathbf{r}_2$. In other word, to become connected, the two operators must have at least one common position index ($\mathbf{r}$). 
This definition can be intuitively viewed as a \emph{connectivity} of the hopping path on the lattice. With this definition, we can regroup the product of the perturbations as follows
\begin{align}
\mathcal{H}'(\tau_2)\mathcal{H}'(\tau_1)=\underbrace{\mathcal{H}'\!\!\!\!\!\overbracket{\,\,\,\,\,(\tau_2)\mathcal{H}'\!\!\!}\,\,\,(\tau_1)}_{\text{connected}}+\underbrace{\overbracket{\mathcal{H}'}(\tau_2)\overbracket{\mathcal{H}'}(\tau_1)}_{\text{disconnected}}.
\end{align}
In the first term, it only contains those products in which $\mathcal{O}^{(\tau_1)}$ and $\mathcal{O}^{(\tau_2)}$ are \textit{connected}
\begin{align}
\mathcal{H}'\!\!\!\!\!\!\overbracket{\,\,\,\,{(\tau_2)}\mathcal{H}'\!\!\!}\,\,\,{(\tau_1)}&=\sum_{\mathbf{r}}\Big[\sum_{\mathbf{r}'}(\mathcal{O}_{\mathbf{r}\mathbf{r}'}^{(\tau_2)}\mathcal{O}_{\mathbf{r}\mathbf{r}'}^{(\tau_1)}+\mathcal{O}_{\mathbf{r}\mathbf{r}'}^{(\tau_2)}\mathcal{O}_{\mathbf{r}'\mathbf{r}}^{(\tau_1)})\notag\\
&\!+\!\!\!\sum_{\mathbf{r}_1\neq\mathbf{r}_2'}\mathcal{O}_{\mathbf{r}\mathbf{r}'_2}^{(\tau_2)}\mathcal{O}_{\mathbf{r}_1\mathbf{r}}^{(\tau_1)}
\!+\!\!\!\sum_{\mathbf{r}_1\neq\mathbf{r}_2}\mathcal{O}_{\mathbf{r}_2\mathbf{r}}^{(\tau_2)}\mathcal{O}_{\mathbf{r}_1\mathbf{r}}^{(\tau_1)}\notag\\
&\!+\!\!\!\sum_{\mathbf{r}_1'\neq\mathbf{r}_2}\mathcal{O}_{\mathbf{r}_2\mathbf{r}}^{(\tau_2)}\mathcal{O}_{\mathbf{r}\mathbf{r}_1'}^{(\tau_1)}
\!+\!\!\!\sum_{\mathbf{r}_1'\neq\mathbf{r}_2'}\mathcal{O}_{\mathbf{r}\mathbf{r}_2'}^{(\tau_2)}\mathcal{O}_{\mathbf{r}\mathbf{r}_1'}^{(\tau_1)}\Big]\label{eqn:conn2}.
\end{align}
In the second term, the operators $\mathcal{O}$ in each products are \textit{disconnected} to each other
\begin{align}\label{eqn:disconn2}
\overbracket{\mathcal{H}'}(\tau_2)\overbracket{\mathcal{H}'}(\tau_1)\!=\!\sum_{\mathbf{r}_1\neq \mathbf{r}_2\mathbf{r}_2'}\sum_{\mathbf{r}_1'\neq \mathbf{r}_2,\mathbf{r}_2'}\mathcal{O}_{\mathbf{r}_2\mathbf{r}_2'}(\tau_2)\mathcal{O}_{\mathbf{r}_1\mathbf{r}_1'}(\tau_1).
\end{align}
Here, the contraction notation $\mathcal{H}'\!\!\!\!\!\overbracket{\,\,\,\,\,(\tau_2)\mathcal{H}'\!\!\!}\,\,\,(\tau_1)$ denotes the collection of all the connected terms in $\mathcal{H}'(\tau_2)\mathcal{H}'(\tau_1)$. 

Using Eq. \eqref{eqn:conn2}, we then calculate $\mu_2$. This leads to
\begin{align*}
\mu_2\!=\!&\int^\beta_0\!\!d \tau_2 d\tau_1\langle \mathcal{T}[\mathcal{H}'\!\!\!\!\!\overbracket{\,\,\,\,\,(\tau_2)\mathcal{H}'\!\!\!}\,\,\,(\tau_1)\!+\!\overbracket{\mathcal{H}'}(\tau_2)\overbracket{\mathcal{H}'}(\tau_1)]\rangle
\!=\!C_2.
\end{align*}
In the above, we note that the second and third line in Eq. \eqref{eqn:conn2} and Eq. \eqref{eqn:disconn2} correspond to the hopping processes with an \textit{open-path}. These open-path connected terms have zero expectation value since their hopping processes end up with an excited state as the final state. These processes are always projected out by the initial ground state. Therefore, the only nonzero contribution is in the first line of Eq. \eqref{eqn:conn2} and the hopping processes of these connected graphs form a \textit{closed-path} on the lattice.

Similarly, we generalize the definition to the case with $n>2$ in Eq. \eqref{eqn:prod}. For $\prod^n_{j=1}\mathcal{O}_{\mathbf{r}_j\mathbf{r}_j'}(\tau_j)$ with any integer $n>1$ is said to be a $n$-\textit{connected graph}, if the product of $\mathcal{O}$s cannot be partitioned into two group such that all the $\mathcal{O}$s in one group have no common position index ($\mathbf{r}$) with the $\mathcal{O}$s in the other group. Similarly, we denote $\overbracket{\!\!\mathcal{H}'(\tau_n)\;\;\;\,}\!\!\!\!\!\!\cdots\!\!\!\!\overbracket{\;\;\;\mathcal{H}'\!\!}\;(\tau_1)$ as the collection of all the $n$-connected graphs. 

We proceed to $\mu_4$ and regroup $\prod_{n=1}^4\mathcal{H}'(\tau_n)$ according to the connectivity in the product of $\mathcal{O}$s. Keeping only the fully contracted terms (closed-path hopping), we have
\begin{align}\label{eqn:mu_4neq0}
\mu_4=&
\int_0^\beta d \tau_1\dots d \tau_4\Big[\langle\mathcal{T}\mathcal{H}'\!\!\!\!\overbracket{\,\,\,(\tau_4)\mathcal{H}'\!\!}\!\!\overbracket{\,\,\,(\tau_3)\mathcal{H}'\!\!}\!\!\overbracket{\,\,\,(\tau_2)\mathcal{H}'\!\!\!}\,\,\,(\tau_1)\rangle\notag\\
&+\langle\mathcal{T}\mathcal{H}'\!\!\!\!\overbracket{\,\,\,(\tau_4)\mathcal{H}'\!\!\!}\,\,(\tau_3)\mathcal{H}'\!\!\!\!\overbracket{\,\,\,(\tau_2)\mathcal{H}'\!\!\!}\,\,(\tau_1)\rangle\notag\\
&+\langle\mathcal{T}\mathcal{H}'\!\!\!\!\overbracket{\,\,\,(\tau_4)\mathcal{H}'\!\!\!\!\underbracket{\,\,\,(\tau_3)\mathcal{H}'\!\!\!}\,\,\,(\tau_2)\mathcal{H}'\!\!\!}\,\,\,(\tau_1)\rangle\notag\\
&
+\langle\mathcal{T}\mathcal{H}'\rlap{$\!\!\!\!\overbracket{\,\,\,(\tau_4)\mathcal{H}'(\tau_3)\mathcal{H}'\!\!\!}\,\,\,$}\;\quad\quad\!\!\!\!\underbracket{\,\,\,\,\;\quad\quad\quad(\tau_2)\mathcal{H}'\!\!\!}\,\,\,(\tau_1)\rangle\Big].
\end{align}
The above last three terms are the disconnected graphs which are the products of two $2$-connected graphs $C_2=\mu_2$. This leads to
\begin{align*}
\mu_4=&\int_0^\beta d \tau_1\dots d \tau_4\langle
\mathcal{T}\mathcal{H}'\!\!\!\!\overbracket{\,\,\,(\tau_4)\mathcal{H}'\!\!}\!\!\overbracket{\,\,\,(\tau_3)\mathcal{H}'\!\!}\!\!\overbracket{\,\,\,(\tau_2)\mathcal{H}'\!\!\!}\,\,\,(\tau_1)\rangle
+3\mu^2_2
\end{align*}
We can see that the disconnected graphs (second terms in the above) are canceled out exactly in $C_4$ in Eq. \eqref{eqn:Gc}. This is guaranteed by the linked-cluster theorem\cite{Metzner:PRB43(1991)}.

Similarly, one can carry out the proof for $n=6$ by rewriting
\begin{align}
C_6
=&\mu_6-15C_4C_2-15C_2^3.\label{eqn:C6-disconn}
\end{align}
In $\mu_6$, there are $\binom{6}{2}=15$ possible contractions that give $C_4C_2$. This can be obtained by first contracting any pair of $\mathcal{O}s$ to form $C_2$ and then contracting the rest of $\mathcal{H}'$s to form $C_4$. To obtain all possible contractions that lead to $C_2^3$, we first contract $\mathcal{H}'(\tau_1)$ with the others. This gives five possible pair contractions. Then, we perform pair contraction for the rest of $\mathcal{H}'$. This has three possibilities. Therefore, there are $5\times3=15$ $C_2^3$ terms in $\mu_6$. All these disconnected terms are again canceled out in Eq. \eqref{eqn:C6-disconn}. Following a similar calculation, one can verify the linked-cluster theorem beyond $n=6$.

Introducing the notion of the connected graph and the linked cluster theorem not only simplifies the calculation of $C_n$ but is also essential for validating the perturbative calculation. To see this, we note that each connected graph scales linearly as $\beta$ in low-temperature. Therefore, the scaling of a disconnected graph is $\beta^{n}$ with $n\geq2$ since the disconnected graphs are a product of connected graphs. If the exact cancellation of the disconnected graphs in $C_n$ is no guarantee, the exponent in Eq. \eqref{eqn:connect} may scale as $\beta^{n}$. This is problematic for the perturbative expansion in Eq. \eqref{eqn:connect}, since in the zero temperature limit, $\beta\to\infty$, the higher-order perturbations become the more and more relevant.

\section{imaginary-time evolution of the excited states}\label{app:states}

The connected correlation function $C_n$ can be obtained by calculating $\prod_{j=1}^n\mathcal{H}'(\tau_j)|\tilde{\Psi}\rangle$ where $\mathcal{H}'(\tau_j)=\mathrm{e}^{\tau_j \mathcal{H}_0}\mathcal{H}'\mathrm{e}^{-\tau_j \mathcal{H}_0}$. Before doing this, we need to specify the choice of the coordinate system that quantizes the spin, since the spin quantization axis cannot be chosen arbitrarily due to the spin-orbit coupling. In the TB model, the spin-orbit coupling Hamiltonian may be written as\cite{Jaffe:SSComm62(1987),Konschuh:PRB82(2010)} Eq.\eqref{eqn:Vlambda}, where the orbital and spin angular momentum are quantized with the \emph{same} local $\bar{x}\bar{y}\bar{z}$-coordinate system (see Fig. \ref{fig:1L-CrX3}). Although using these local coordinate systems gives a simple expression to Eq.\eqref{eqn:Vlambda}, the cost of doing so is that the local descriptions of the spin become incompatible between different X ions across the lattice. In order to describe the spin with the same reference, we transform each of these local bases to a new basis where all the spins are quantized in the same global $xyz$-system (Fig.\ref{fig:1L-CrX3}). This can be achieved by rotating the spin with the unitary matrix $\mathcal{S}^{\mathbf{r}}_{\sigma\bar{\sigma}}$ as $p_{\mathbf{r}\alpha \sigma }=\sum_{\bar{\sigma}}\mathcal{S}^{\mathbf{r}}_{ \sigma\bar{\sigma} } p_{\mathbf{r}\alpha \bar{\sigma} }$. This rotation leads to the transformation of the Pauli matrices as
\begin{equation}\label{eqn:R}
\sum_{\bar{\sigma}_1\bar{\sigma}_2}\mathcal{S}^{\mathbf{r}\dagger}_{\sigma_1 \bar{\sigma}_1}\bar{\tau}^{\alpha}_{\bar{\sigma}_1\bar{\sigma}_2}\mathcal{S}^{m}_{\bar{\sigma}_2\sigma_2}=\sum_{j=x,y,z}\mathcal{R}^{\mathbf{r}}_{j\alpha}\tau^{j}_{\sigma_1\sigma_2}.
\end{equation}
In the above, $\mathcal{R}^{\mathbf{r}}_{j\alpha}$ is an orthogonal rotational matrix that changes the local $\bar{x}\bar{y}\bar{z}$-coordinate system to the global $xyz$-coordinate system. The index $j$ corresponds to the global $x,y,z$ axis instead of the $p$-orbital indices. With this transformation, this leads to
\begin{equation}
\mathcal{V}_\lambda=\sum_{\mathbf{r} \sigma_1 \sigma_2}\sum_{\alpha_1 \alpha_2j}\Lambda_{\mathbf{r},j}^{\alpha_1\alpha_2}\tau^{j}_{\sigma_1\sigma_2} p^{\dagger}_{\mathbf{r}_1\alpha_1 \sigma_1}p_{\mathbf{r}_2\alpha_2 \sigma_2}
\end{equation}
with
$\Lambda_{\mathbf{r},j}^{\alpha_1 \alpha_2}=i\sum_{\alpha}\lambda^\alpha\mathcal{R}^{\mathbf{r}}_{ j \alpha}\varepsilon_{\alpha\alpha_2 \alpha_1}
$.
Therefore, in the new basis, all the spin-quantization axes are defined uniformly across the crystal in the $xyz$ system.

Furthermore, it is preferable to transform the $d$-orbital operator in the hopping Hamiltonian $\mathcal{H}_t$ and $\mathcal{H}_d$ as
\begin{align}
\mathcal{H}^\dagger_t=&\sum_{\dot{1}1}\sum_{\dot{\sigma}_1\sigma_1}  t_{\dot{1}1 }\mathsf{d}^\dagger_{ \dot{1}\dot{\sigma}_1}\bar{\chi}^{\dot{\sigma}_1}_{ \dot{\mathbf{r}}_1\sigma_1} p_{1 \sigma_1},\\
\mathcal{H}_d=&\sum_{\dot{1}\dot{2}}\sum_{\dot{\sigma}_1\dot{\sigma}_2}  u_{\dot{1}\dot{2} }\mathsf{d}^\dagger_{ \dot{1}\dot{\sigma}_1}\mathsf{d}_{ \dot{2}\dot{\sigma}_2} \bar{\chi}^{\dot{\sigma}_1}_{ \dot{\mathbf{r}}_1\sigma}\chi^{\dot{\sigma}_2}_{ \dot{\mathbf{r}}_2\sigma},
\end{align}
where the $d$-orbital spin orientation is parallel or antiparallel with the localized spin in the $t_{2g}$ at Cr site. In the above equation, we have used the compact notation $n\equiv\mathbf{r}_n ,\alpha_n$ and $\dot{n}\equiv\dot{\mathbf{r}}_n ,\dot{\alpha}_n$ to simplify the expression. 

With all these settings, we first calculate
\begin{align}
\mathcal{H}'(\tau_1)&|\tilde{\Psi}\rangle=[\mathcal{H}^\dagger_t(\tau_1)+\mathcal{H}_d(\tau_1)]|\tilde{\Psi}\rangle.
\end{align}
To evaluate the above, we use
\begin{align*}
\mathrm{e}^{\tau\mathcal{H}_0}&\mathsf{d}^\dagger_{\dot{\mathbf{r}}_1 \dot{\alpha}_1\dot{\sigma}_1}|\tilde{\Psi}\rangle
=
\sum_{m=0}^{3}\mathrm{e}^{\tau \Omega^m_{\dot{1}\dot{\sigma}_1}}\Upsilon^{m,\dagger}_{ \dot{\mathbf{r}}_1\dot{\alpha}_1\dot{\sigma}_1}|\tilde{\Psi}\rangle,
\end{align*}
where $\Omega^m_{\dot{1}\dot{\sigma}_1}$ is the one-$d$-electron excitation energy in Eq. \eqref{eqn:Omega}, and
\begin{align}
\Upsilon^{m\dagger}_{ \dot{\mathbf{r}}\dot{\alpha}\dot{\sigma}}= &[\delta_{ \dot{\sigma},-}P_{ \dot{\alpha}_1}\!+\!\delta_{ \dot{\sigma},+}(1-P_{ \dot{\alpha}})]\delta_{m,0}\mathsf{d}^{\dagger}_{ \dot{\mathbf{r}} \dot{\alpha} \dot{\sigma}}\notag\\
&+\delta_{ \dot{\sigma},-}(1-P_{ \dot{\alpha}})A^m_{ \dot{\alpha} }\sum_{k=0}^3 v^{k}_{m,\dot{\alpha}}\phi^{k\dagger}_{ \dot{\mathbf{r}} \dot{\alpha}}
\end{align}
with 
$P_{\dot{\alpha}}=1$ ($P_{\dot{\alpha}}=0$) if $\dot{\alpha}=1,2,3$ ($\dot{\alpha}=4,5$). This yields
\begin{align}
\mathcal{H}'(\tau_1)&|\tilde{\Psi}\rangle
=
[t^{m_1}_{\dot{1}1}(\tau_1)\bar{\chi}^{\dot{\sigma}_1}_{ \dot{\mathbf{r}}_1\sigma_1}\Upsilon^{m_1,\dagger}_{\dot{1} \dot{\sigma}_1}p_{1 \sigma_1}\notag\\
&+ u_{\dot{1}\dot{2} }(\tau_1)\bar{\chi}^{\dot{\sigma}_1}_{ \dot{\mathbf{r}}_1\sigma}\chi^{\dot{\sigma}_2}_{ \dot{\mathbf{r}}_2\sigma}\Upsilon^{m_1,\dagger}_{\dot{1} \dot{\sigma}_1}\mathsf{d}_{ \dot{2}\dot{\sigma}_2} ]|\tilde{\Psi}\rangle,\label{eqn:Ht1}
\end{align}
where we have let $t^m_{\dot{1} 1} (\tau_1)=\mathrm{e}^{\tau_1 \Omega^m_{\dot{1}\dot{\sigma}_1}}t_{\dot{1}   1}\mathrm{e}^{-\tau_1 \bar{\nu}_{ 1}}$ and $u^m_{\dot{1} \dot{2}} (\tau_1)=\mathrm{e}^{\tau_1 \Omega^m_{\dot{1}\dot{\sigma}_1}}u_{\dot{1}   \dot{2}}\mathrm{e}^{-\tau_1 \bar{\omega}_{\dot{2}}}$.

Following the similar procedure, it is straightforward to calculate $\mathcal{H}'(\tau_2)\mathcal{H}'(\tau_1)|\tilde{\Psi}\rangle$. Here, we list the calculated results as follows
\begin{align}
	\mathcal{H}_t(\tau_2)\mathcal{H}^\dagger_t(\tau_1)|\tilde{\Psi}\rangle=&\bar{\chi}^{\dot{\sigma}_1}_{ \dot{\mathbf{r}}_1\sigma}\chi^{+}_{ \dot{\mathbf{r}}_2\sigma}P_{ \dot{\alpha}_2}t^{m_1}_{\dot{1} 1}(\tau_1)  \bar{t}_{ 1 \dot{2} }(\tau_2)\notag\\
&
\mathsf{d}_{\dot{2} ,+}
\Upsilon^{m_1\dagger}_{\dot{1} \dot{\sigma}_1} 
|\tilde{\Psi}\rangle,\label{eqn:HH+}\\
	\mathcal{H}^\dagger_t(\tau_2)\mathcal{H}^\dagger_t(\tau_1)|\tilde{\Psi}\rangle=&\bar{X}^{\eta,\dot{\sigma}_1\dot{\sigma}_2}_{\dot{\mathbf{r}}_1\dot{\mathbf{r}}_2,\sigma_1 \sigma_2}
   t^{m_1}_{\dot{1} 1} (\tau_1)t^{m_2}_{\dot{2} 2} (\tau_2)\mathrm{e}^{\tau_2 \Theta^{\eta }_{12}}\notag\\
&p_{ 2\sigma_2}p_{ 1\sigma_1}
\Upsilon^{m_2 \dagger}_{\dot{2} \dot{\sigma}_2} \Upsilon^{m_1 \dagger}_{\dot{1} \dot{\sigma}_1} 
|\tilde{\Psi}\rangle,\label{eqn:HdHd}
	\\
	\mathcal{V}_\lambda(\tau_2)\mathcal{H}^\dagger_t(\tau_1)|\tilde{\Psi}\rangle=&
 \bar{\chi}^{\dot{\sigma}_1}_{ \dot{\mathbf{r}}_1\sigma_1}t^{m_1}_{\dot{1} 1} (\tau_1)\Lambda_{\mathbf{r}_2j}^{\alpha_1 \alpha_2}(\tau_2)\tau^{j}_{\sigma_1 \sigma_2}\delta_{\mathbf{r}_1\mathbf{r}_2}\notag\\
& p_{ 2 \sigma_2}
\Upsilon^{m_1 \dagger}_{\dot{1} \dot{\sigma}_1}
|\tilde{\Psi}\rangle,
\\
\mathcal{H}_\perp(\tau_2)\mathcal{H}^\dagger_t(\tau_1)|\tilde{\Psi}\rangle=&\bar{\chi}^{\dot{\sigma}_1}_{ \dot{\mathbf{r}}_1\sigma_1}
t^{m_1}_{\dot{1}1}(\tau_1) T^{\ell_1\ell_2}_{12}(\tau_2)\notag\\
&p^{\ell_2}_{2 \sigma_1}\Upsilon^{m_1\ell_1,\dagger}_{\dot{1} \dot{\sigma}_1}|\tilde{\Psi}\rangle,
\end{align}
where we let $\bar{t}_{ 1\dot{1} }(\tau_1)=\mathrm{e}^{\tau_1 \bar{\nu}_{ 1}}t_{ 1 \dot{1} } \mathrm{e}^{-\tau_1 \bar{\omega}_{\dot{1}}}$, $\Lambda_{\mathbf{r}_1j}^{\alpha_1 \alpha_2}(\tau_2)=\mathrm{e}^{\tau_2 \bar{\nu}_{\mathbf{r}_1\alpha_1}}\Lambda_{\mathbf{r}_1j}^{\alpha_1 \alpha_2}\mathrm{e}^{-\tau_2 \bar{\nu}_{\mathbf{r}_1\alpha_2}}$, and $\Theta^{\eta }_{12}=\Theta^{\eta,\alpha_1\alpha_2 }_{\mathbf{r}_1\mathbf{r}_2}$.
In Eq. \eqref{eqn:HdHd}, we let the spin-singlet and spin-triplet pair in the $d$ orbital as
\begin{equation*}
\bar{X}^{\eta,\dot{\sigma}_1\dot{\sigma}_2}_{\dot{\mathbf{r}}_1\dot{\mathbf{r}}_2,\sigma_1 \sigma_2}
=
\frac{1}{2}(\bar{\chi}^{\dot{\sigma}_1}_{\dot{\mathbf{r}}_1\sigma_1}\bar{\chi}^{\dot{\sigma}_2}_{ \dot{\mathbf{r}}_2\sigma_2}+\eta\bar{\chi}^{\dot{\sigma}_1}_{ \dot{\mathbf{r}}_1\sigma_2}\bar{\chi}^{\dot{\sigma}_2}_{ \dot{\mathbf{r}}_2\sigma_1}).
\end{equation*}

We can further calculate the higher order excitations that involve the spin-flip processes from the spin-orbit coupling
\begin{widetext}
\begin{align}
\mathcal{V}_\lambda(\tau_3)\mathcal{H}^\dagger_t(\tau_2)\mathcal{H}^\dagger_t(\tau_1)|\tilde{\Psi}\rangle\!
=&
-\frac{1}{2}[ t^{m_1}_{\dot{1}  1} (\tau_1) t^{m_2}_{\dot{2} 2}(\tau_2) -\eta t^{m_1}_{\dot{1}  1} (\tau_1) t^{m_2}_{\dot{2} 2}(\tau_2) ]\mathrm{e}^{(\tau_2-\tau_3)\Theta^{\eta }_{12}}\Lambda_{\mathbf{r}_3j}^{\alpha_2\alpha_3}(\tau_3)\mathrm{e}^{\tau_3\Theta^{\eta'}_{13}}
\notag\\
&
(\bar{X}^{\eta,\dot{\sigma}_1\dot{\sigma}_2}_{\dot{\mathbf{r}}_1\dot{\mathbf{r}}_2,\sigma_1\sigma_2}\tau^{j}_{\sigma_2\sigma_3}+\eta'\bar{X}^{\eta,\dot{\sigma}_1\dot{\sigma}_2}_{\dot{\mathbf{r}}_1\dot{\mathbf{r}}_2,\sigma_3\sigma_2}\tau^{j}_{\sigma_2\sigma_1})\delta_{\mathbf{r}_2\mathbf{r}_3}p_{ 3\sigma_3}p_{ 1\sigma_1}
\Upsilon^{m_2 \dagger}_{\dot{2} \dot{\sigma}_2}\Upsilon^{m_1 \dagger}_{\dot{1} \dot{\sigma}_1}|\tilde{\Psi}\rangle,
\\
\mathcal{H}_{t}(\tau_3)\mathcal{V}_\lambda(\tau_2)\mathcal{H}^\dagger_t(\tau_1)|\tilde{\Psi}\rangle\!
=&-P_{ \dot{\alpha}_3}   t^{m_1}_{\dot{1}  1} (\tau_1)\Lambda_{\mathbf{r}_2j}^{\alpha_1 \alpha_2}(\tau_2)\bar{t}_{2\dot{3}}(\tau_3)\delta_{\mathbf{r}_1\mathbf{r}_2}\chi^{+}_{ \dot{\mathbf{r}}_3\sigma_2}\tau^{j}_{\sigma_1 \sigma_2}\bar{\chi}^{\dot{\sigma}_1}_{ \dot{\mathbf{r}}_1\sigma_1}\mathsf{d}_{\dot{\mathbf{r}}_3 \dot{\alpha}_3,+}\Upsilon^{m_1 \dagger}_{\dot{1} \dot{\sigma}_1} |\tilde{\Psi}\rangle,
\\
\mathcal{H}^\dagger_t(\tau_3)\mathcal{V}_\lambda(\tau_2)\mathcal{H}^\dagger_t(\tau_1)|\tilde{\Psi}\rangle\!
=& t^{m_1}_{\dot{1}  1} (\tau_1) \Lambda_{\mathbf{r}_2j}^{\alpha_1 \alpha_2}(\tau_2)t^{m_3}_{ \dot{3}3}(\tau_3)\frac{1}{2}(\bar{\chi}^{\dot{\sigma}_1}_{ \dot{\mathbf{r}}_1\sigma_1}\tau^{j}_{\sigma_1 \sigma_2}\bar{\chi}^{\dot{\sigma}_3}_{\dot{\mathbf{r}}_3\sigma_3}+\eta\bar{\chi}^{\dot{\sigma}_1}_{ \dot{\mathbf{r}}_1\sigma_1}\tau^{j}_{\sigma_1 \sigma_3}\bar{\chi}^{\dot{\sigma}_3}_{\dot{\mathbf{r}}_3\sigma_2})\notag\\
&\mathrm{e}^{\tau_3\Theta^{\eta}_{23}}\delta_{\mathbf{r}_2\mathbf{r}_1} p_{ 3\sigma_3}p_{ 2\sigma_2}\Upsilon^{m_3 \dagger}_{\dot{3} \dot{\sigma}_3}\Upsilon^{m_1 \dagger}_{\dot{1} \dot{\sigma}_1}|\tilde{\Psi}\rangle,
\end{align}
and the excitations created by inter-layer hopping
\begin{align}
\mathcal{H}_{\perp}(\tau_3)\mathcal{H}^{\dagger}_{t}(\tau_2)\mathcal{H}^{\dagger}_{t}(\tau_1)|\tilde{\Psi}\rangle\!=
	& 
\bar{X}^{\eta,\ell_1\dot{\sigma}_1\ell_2\dot{\sigma}_2}_{\dot{\mathbf{r}}_1\ell_1\dot{\mathbf{r}}_2,\sigma_1 \sigma_2}
\Big[
 t^{m_2}_{\dot{2} 2}(\tau_1)t^{m_1}_{\dot{1}  1} (\tau_2)+t^{m_1}_{\dot{1}  1} (\tau_1)t^{m_2}_{\dot{2} 2}(\tau_2)\Big]
T^{\ell_1\ell_2}_{ 13}(\tau_3)\mathrm{e}^{\tau_3\Theta^{\eta }_{23}}\notag\\
&
p^{\ell_2}_{ 3\sigma_1}p^{\ell_2}_{ 2\sigma_2}
\Upsilon^{m_2\ell_2\dagger}_{ \dot{2}\dot{\sigma}_2}\Upsilon^{m_1\ell_1\dagger}_{\dot{1}\dot{\sigma}_1}|\tilde{\Psi}\rangle,\\
	\mathcal{H}_{t}(\tau_3)\mathcal{H}_{\perp}(\tau_2)\mathcal{H}^{\dagger}_{t}(\tau_1)|\tilde{\Psi}\rangle\!=&
 -t^{m_1}_{\dot{1}  1} (\tau_1)T^{\ell_1\ell_2}_{ 1 2}(\tau_2)  \bar{t}_{ 2 \dot{ 2}}(\tau_3)\bar{\chi}^{\ell_1,\dot{\sigma}_1}_{\dot{\mathbf{r}}_1\sigma_1}\chi^{\ell_2,+}_{\dot{\mathbf{r}}_2\sigma_1}\mathsf{d}^{\ell_2}_{ \dot{2},+} \Upsilon^{m_1\ell_1\dagger}_{\dot{1}\dot{\sigma}_1}|\tilde{\Psi}\rangle,\\
	\mathcal{H}^\dagger_t(\tau_3)\mathcal{H}_{\perp}(\tau_2)\mathcal{H}^\dagger_t(\tau_1)|\tilde{\Psi}\rangle\!=&
\eta\bar{X}^{\eta,\ell_1\dot{\sigma}_1\ell_2\dot{\sigma}_2}_{\dot{\mathbf{r}}_1\ell_1\dot{\mathbf{r}}_2,\sigma_1 \sigma_2}
t_{\dot{1} 1}^{ m_1}(\tau_1)T^{\ell_1\ell_2}_{1 2}(\tau_2) t^{m_2}_{ \dot{2} 3}(\tau_3) \mathrm{e}^{\tau_3\Theta^{\eta}_{23}}
p^{\ell_2}_{ 3 \sigma_1}p^{\ell_2}_{ 2 \sigma_2} \Upsilon^{m_2\ell_2\dagger}_{ \dot{2}\dot{\sigma}_2}\Upsilon^{m_1\ell_1\dagger}_{\dot{ 1}\dot{\sigma}_1}|\tilde{\Psi}\rangle,\label{eqn:H+H''H+}
\end{align}
\end{widetext}
where $T^{\ell_1\ell_2}_{1 2}(\tau)=\mathrm{e}^{\tau\bar{\nu}_1}T^{\ell_1\ell_2}_{1 2}\mathrm{e}^{-\tau\bar{\nu}_2} $ and
\begin{equation*}
\bar{X}^{\eta,\ell_1\dot{\sigma}_1,\ell_2\dot{\sigma}_2}_{\dot{\mathbf{r}}_1\dot{\mathbf{r}}_2,\sigma_1 \sigma_2}
=
\frac{1}{2}(\bar{\chi}^{\ell_1,\dot{\sigma}_1}_{\dot{\mathbf{r}}_1\sigma_1}\bar{\chi}^{\ell_2,\dot{\sigma}_2}_{ \dot{\mathbf{r}}_2\sigma_2}+\eta\bar{\chi}^{\ell_1,\dot{\sigma}_1}_{ \dot{\mathbf{r}}_1\sigma_2}\bar{\chi}^{\ell_2,\dot{\sigma}_2}_{ \dot{\mathbf{r}}_2\sigma_1}).
\end{equation*}
After having the $\tau$-evolution of the excited states in Eqs. \eqref{eqn:HH+} to \eqref{eqn:H+H''H+}, we can proceed to
the calculation of $C_n$ by using 
\begin{equation*}
\langle\tilde{\Psi}|\mathcal{H}'(\tau_n)\dots\mathcal{H}'(\tau_1)\!=\!\Big[\mathcal{H}'(-\tau_1)\dots\mathcal{H}'(-\tau_n)|\tilde{\Psi}\rangle\Big]^\dagger.
\end{equation*}
To do this calculation, we first introduce the following notation
\begin{equation}
G_n(\tau_n\dots\tau_1)=\langle\tilde{\Psi}|\mathcal{H}'(\tau_n)\dots\mathcal{H}'(\tau_1)|\tilde{\Psi}\rangle.\label{eqn:Gn}
\end{equation}
Therefore, we have $G_2=\langle\tilde{\Psi}|\mathcal{H}'(\tau_2)\mathcal{H}'(\tau_1)|\tilde{\Psi}\rangle$
\begin{align*}
G_2(\tau_2\tau_1)=&\Gamma^{m_1}_{ \dot{\alpha}_1 \dot{\sigma}_1}[\mathrm{e}^{(\tau_1-\tau_2)(\Omega^{m_1}_{\dot{\mathbf{r}}_1\dot{\alpha}_1\dot{\sigma}_1}-\bar{\nu}_{\mathbf{r}_1\alpha_1})} t_{1 \dot{1}}   t_{\dot{1} 1 }\notag\\
&+
\mathrm{e}^{(\tau_1-\tau_2)(\Omega^{m_1}_{\dot{\mathbf{r}}_1\dot{\alpha}_1\dot{\sigma}_1}-\bar{\omega}_{\dot{2}})}
 u_{\dot{1} \dot{2}}   u_{\dot{2} \dot{1} }].
\end{align*}
In the above calculation, we have used the following commutation relations
\begin{align}
\langle\tilde{\Psi}|\{\Upsilon^{m_2}_{ \dot{2} \dot{\sigma}_2},&\Upsilon^{ m_1\dagger}_{\dot{1} \dot{\sigma}_1}\}|\tilde{\Psi}\rangle=\Gamma^{m_1}_{ \dot{\alpha}_1 \dot{\sigma}_1}\delta(\dot{1}\dot{2})
\label{eqn:dd}
\end{align}
with $\delta(\dot{1}\dot{2})=\delta_{\dot{\mathbf{r}}_1\dot{\mathbf{r}}_2}\delta_{m_1m_2}\delta_{\dot{\alpha}_1\dot{\alpha}_2}\delta_{\dot{\sigma}_1\dot{\sigma}_2}$ and the factor $\Gamma^m_{ \dot{\alpha}_1\dot{\sigma}_1}$ is given by \eqref{eqn:Gamma}.
Therefore, using $G_2(\tau_2\tau_1)$ and integrating out $\tau$, we obtain $C_2$ in the $\beta\to\infty$ limit as 
\begin{align}
C_2
=&2!\beta\Big[\frac{t_{ 1 \dot{1}}  t_{\dot{1} 1 }}{\mathcal{E}^{m_1}_{\dot{1}1}}+\delta_{\dot{\sigma}_2,+}|\bar{\chi}^{\dot{\sigma}_1}_{\dot{\mathbf{r}}_1\sigma}\chi^{\dot{\sigma}_2}_{\dot{\mathbf{r}}_2\sigma}|^2\frac{P_{\dot{\alpha}_2}u_{\dot{1}\dot{2}}u_{\dot{2}\dot{1}}}{\Omega^{m_1}_{ \dot{1}\dot{\sigma}_1}-\bar{\omega}_{ \dot{2}}}\Big].
\end{align}
where the sum of all the indices in the right-hand-side is implicitly assumed. In the above, the first term corresponds to the spin-independent energy correction to the ground state. It is irrelevant to the magnetic states. We discard this correction in the paper. 

\begin{table}
\caption{\label{tbl:spin-sum}The the spin-sum of  the spin wavefunction $\bm{\chi}_{\dot{\mathbf{r}}\dot{\sigma}}^\dagger=[\bar{\chi}^{\dot{\sigma}}_{ \dot{\mathbf{r}}\uparrow}, \bar{\chi}^{\dot{\sigma}}_{ \dot{\mathbf{r}}\downarrow}]$, and $\bm{\tau}=[\tau^x,\tau^y,\tau^z]$ are Pauli's matrices.}
\begin{ruledtabular}
\begin{tabular}{ll}
spin products & spin-sum result \\
\hline
$\bm{\chi}_{\dot{\mathbf{r}}\dot{\sigma}}^\dagger\bm{\tau}\bm{\chi}_{\dot{\mathbf{r}}\dot{\sigma}}$&$\dot{\sigma}\mathbf{s}_{ \dot{\mathbf{r}}}$\\
$|\bm{\chi}_{\dot{\mathbf{r}_2}\dot{\sigma}_2}^\dagger\bm{\chi}_{\dot{\mathbf{r}_1}\dot{\sigma}_1}|^2$
& $\frac{1}{2}(1+\dot{\sigma}_1\dot{\sigma}_2\mathbf{s}_{ \dot{\mathbf{r}}_2}\cdot\mathbf{s}_{ \dot{\mathbf{r}}_1})$
\\
\hline
\multirow{2}*{$\bm{\chi}_{\dot{\mathbf{r}_1}\dot{\sigma}_1}^\dagger\bm{\tau}\bm{\chi}_{\dot{\mathbf{r}_2}\dot{\sigma}_2}\bm{\chi}_{\dot{\mathbf{r}_2}\dot{\sigma}_2}^\dagger\bm{\chi}_{\dot{\mathbf{r}_1}\dot{\sigma}_1}$ } 
& $\frac{1}{2}[-\dot{\sigma}_1\mathbf{s}_{ \dot{\mathbf{r}}_1}-\dot{\sigma}_2\mathbf{s}_{ \dot{\mathbf{r}}_2}$ \\
& $\quad+i\dot{\sigma}_1\dot{\sigma}_2\mathbf{s}_{ \dot{\mathbf{r}}_1}\times\mathbf{s}_{ \dot{\mathbf{r}}_2}]$\\
\hline
\multirow{3}*{$\bm{\chi}_{\dot{\mathbf{r}_1}\dot{\sigma}_1}^\dagger\tau^j\bm{\chi}_{\dot{\mathbf{r}_2}\dot{\sigma}_2}\bm{\chi}_{\dot{\mathbf{r}_2}\dot{\sigma}_2}^\dagger\tau^{j'}\bm{\chi}_{\dot{\mathbf{r}_1}\dot{\sigma}_1}$ }
&$\frac{1}{2}[i\varepsilon_{j j' j''}(\dot{\sigma}_1s^{j''}_{ \dot{\mathbf{r}}_1}-\dot{\sigma}_2s^{j''}_{ \dot{\mathbf{r}}_2})$\\
&$\quad+\delta_{j j'}(1-\dot{\sigma}_1\dot{\sigma}_2\mathbf{s}_{ \dot{\mathbf{r}}_2}\cdot\mathbf{s}_{ \dot{\mathbf{r}}_1})$\\
&$\quad+\dot{\sigma}_1\dot{\sigma}_2(s^{j}_{ \dot{\mathbf{r}}_2}s^{j'}_{ \dot{\mathbf{r}}_1}+s^{j'}_{ \dot{\mathbf{r}}_2}s^j_{ \dot{\mathbf{r}}_1})]$
\end{tabular}
\end{ruledtabular}
\end{table}
\begin{table*}
\caption{\label{tbl:TBpd}The \textit{ab initio} TB constants for the nearest-neigbhor Cr-X hopping: $t_{\dot{\mathbf{r}}\dot{\alpha},\mathbf{r}\alpha}$ (eV). We note that these TB constants are obtained after performing the rotation in Eq. \eqref{eqn:U-trans}. We only present the hopping between sublattices $A,B$ and to sublattices $h_{1,2}$ (see Fig.\ref{fig:1L-CrX3}). These TB constants are sufficient for calculating all the nearest-neighbor exchange couplings.}
\begin{ruledtabular}
\begin{tabular}{c|ccccccccccccccc}
 Hopping &          & $A\to h_1$&         &&            & $A\to h_2$&         &&            &$B\to h_1$&          &&            &$B\to h_2$& \\
\hline
  $t_{\dot{\mathbf{r}}\dot{\alpha},\mathbf{r}\alpha}$ &$ p_z$&$p_x$&$p_y$&& $ p_z$&$p_x$&$p_y$&& $ p_z$&$p_x$&$p_y$&& $ p_z$&$p_x$&$p_y$\\
\hline
$\dot{\alpha}=1$ &-0.201 & 0.2151 & 0.265 & &-0.2006 & -0.2154 & -0.2692 &&0.2007 & 0.2154 & 0.2692&&0.201 & -0.2151 & -0.265\\
$\dot{\alpha}=2$& -0.1161 & 0.1385 & 0.2185  & &-0.0764 & -0.1809 & 0.43&&0.0767 & 0.1814 & -0.43 &&0.1163 & -0.139 & -0.2178 \\
$\dot{\alpha}=3$& -0.1528 & 0.2842 & -0.3732 & &-0.1745 & -0.2605 & -0.0037 &&0.1743 & 0.2602 & 0.0044&&0.1526 & -0.2839 & 0.3735\\
$\dot{\alpha}=4$& -0.8352 & -0.7048 & -0.0267 & &0.4257 & -0.3715 & -0.0065&&-0.426 & 0.3717 & 0.0066&&0.8352 & 0.7048 & 0.0267 \\
$\dot{\alpha}=5$& 0.007 & 0.0216 & -0.008 & &-0.7195 & 0.5972 & 0.0267&&0.7195 & -0.5971 & -0.0267 &&-0.0067 & -0.0214 & 0.008\\
\end{tabular}
\end{ruledtabular}
\end{table*}
\begin{table}
\caption{\label{tbl:TBdd}The \textit{ab initio} TB constants for the nearest-neigbhor Cr-Cr direct hopping: $u_{\dot{\mathbf{r}}\dot{\alpha},\dot{\mathbf{r}}'\dot{\alpha}'}$ [eV]. We note that these TB constants are obtained after performing the rotation in Eq. \eqref{eqn:W-trans}.}
\begin{ruledtabular}
\begin{tabular}{c|ccccc}
 Hopping &          & &    $A\to B$     &&              \\
\hline
  $u_{\dot{\mathbf{r}}\dot{\alpha},\dot{\mathbf{r}}'\dot{\alpha}'}$ &$\dot{\alpha}=1$&$\dot{\alpha}=2$&$\dot{\alpha}=3$&$\dot{\alpha}=4$ &$\dot{\alpha}=5$\\
\hline
$\dot{\alpha}'=1$ &-0.0273 & -0.0316 & 0.0154 & -0.0023 & 0.053\\
$\dot{\alpha}'=2$& -0.0317 & 0.0133 & 0.0137 & -0.0074 & 0.0363 \\
$\dot{\alpha}'=3$&0.0154 & 0.0137 & -0.01 & 0.0519 & -0.0007\\
$\dot{\alpha}'=4$&-0.0024 & -0.0074 & 0.0518 & -0.0781 & -0.0197 \\
$\dot{\alpha}'=5$&0.053 & 0.0363 & -0.0006 & -0.0199 & -0.0347\\
\end{tabular}
\end{ruledtabular}
\end{table}

To calculate $G_n$ with $n\geq4$, we use the following relation 
\begin{align*}
\langle\tilde{\Psi}|p^\dagger_{4\sigma_4}p^\dagger_{3\sigma_3}&p_{2\sigma_2}p_{1\sigma_1}|\tilde{\Psi}\rangle=\delta(14)\delta(23)-\delta(13)\delta(24)
\end{align*}
with $\delta(12)=\delta_{\mathbf{r}_1\mathbf{r}_2}\delta_{\alpha_1\alpha_2}\delta_{\sigma_1\sigma_2}$, and
\begin{align*}
\langle\tilde{\Psi}|\Upsilon^{m_4}_{ \dot{4}\dot{\sigma}_4}\Upsilon^{m_3}_{\dot{3}\dot{\sigma}_3}&\Upsilon^{ m_2\dagger}_{\dot{2} \dot{\sigma}_2}\Upsilon^{m_1 \dagger}_{\dot{1} \dot{\sigma}_1}|\tilde{\Psi}\rangle\notag\\
&=\Gamma^{m_1}_{ \dot{\alpha}_1\dot{\sigma}_1}\Gamma^{m_2}_{ \dot{\alpha}_2\dot{\sigma}_2}\Big[\delta(\dot{3}\dot{2})\delta(\dot{4}\dot{1})-\delta(\dot{3}\dot{1})\delta(\dot{4}\dot{2})
\Big].
\end{align*}
To get rid of the time-ordering operator, we write the $\tau$-integration as 
\begin{equation}
C_n=(-1)^nn!\int d(n\dots1)G_{n}(\tau_n\dots \tau_1),
\end{equation}
where $\int d(n\dots 21)=\int_0^\beta \!\!d \tau_n\int^{\tau_{n}}_{0}\!\!d \tau_{n-1}\dots\int^{\tau_{2}}_{0}\!\! d \tau_1$ with $0<\tau_1 < \dots < \tau_{n} <\beta $. 

Therefore, we can write $C_n$ as
\begin{align}\label{eqn:C4}
C_4=&4!\int \!\!d(4321)(G^{hh^\dagger hh^\dagger}_{4}\!+G^{hh h^\dagger h^\dagger}_{4}\!+G^{hvv h^\dagger}_{4}),
\end{align}
where 
\begin{align*}
&G^{hh^\dagger hh^\dagger}_{4}\!\!=\!\langle\tilde{\Psi}|\mathcal{H}_t(\tau_4)\mathcal{H}^\dagger_t(\tau_3)\mathcal{H}_t(\tau_2)\mathcal{H}^\dagger_t(\tau_1)|\tilde{\Psi}\rangle_{c},\\
&G^{hh h^\dagger h^\dagger}_{4}\!\!=\!\langle\tilde{\Psi}|\mathcal{H}_t(\tau_4)\mathcal{H}_t(\tau_3)\mathcal{H}^\dagger_t(\tau_2)\mathcal{H}^\dagger_t(\tau_1)|\tilde{\Psi}\rangle_c,\\
&G^{hvv h^\dagger}_{4}\!=\!\langle\tilde{\Psi}|\mathcal{H}_t(\tau_4)\mathcal{V}_\lambda(\tau_3)\mathcal{V}_\lambda(\tau_2)\mathcal{H}^\dagger_t(\tau_1)|\tilde{\Psi}\rangle_c.
\end{align*}
Instead of using contraction notation, we use the subscript $c$ to indicate that only the connected graphs in the expectation values are evaluated. In the superscript of $G_{4}$ in Eq. \eqref{eqn:C4}, $h$, $h^\dagger$, and $v$ represent $\mathcal{H}_t(\tau)$, $\mathcal{H}_t^\dagger(\tau)$, and $\mathcal{V}_\lambda(\tau)$, respectively. The superscript also shows explicitly the ordering of these perturbative Hamiltonians in the product. Similarly, we can write down $C_5$ and $C_6$ by using the same notation
\begin{align}
C_5=&-5!\int \!\!d(5\dots1)(G^{hh^\dagger hvh^\dagger}_{5}\!+G^{hvh^\dagger hh^\dagger}_{5}\!\notag\\
&+G^{hh v h^\dagger h^\dagger}_{5}\!+G^{hhh^\dagger v h^\dagger}_{5}+G^{hvhh^\dagger  h^\dagger}_{5}),\label{eqn:C5}\\
C_6=&6!\int \!\!d(6\dots1)(G^{hh^\dagger hvvh^\dagger}_{6}\!+G^{hvh^\dagger hvh^\dagger}_{6}+G^{hvvh^\dagger hh^\dagger}_{6}\!
\notag\\
&+G^{hh h^\dagger vv h^\dagger}_{6}\!+G^{hvh h^\dagger v h^\dagger}_{6}++G^{hvvh h^\dagger  h^\dagger}_{4}\!\!\notag\\
&+G^{hhvvh^\dagger  h^\dagger}_{6}+G^{hhvh^\dagger v h^\dagger}_{6}+G^{hvhvh^\dagger  h^\dagger}_{6}).\label{eqn:C6}
\end{align}
Furthermore, we can also calculate the connected correlation function for inter-layer exchange couplings
\begin{align}\label{eqn:C6'}
C^{\perp}_6=&6!\int \!\!d(6\dots1)(G^{hTh^\dagger hTh^\dagger}_{6}+G^{hTh h^\dagger T h^\dagger}_{6}\notag\\
&+G^{hhTTh^\dagger  h^\dagger}_{6}+G^{hhTh^\dagger T h^\dagger}_{6}+G^{hThTh^\dagger  h^\dagger}_{6}).
\end{align}
The superscript $T$ in $G_{6}$ stands for inter-layer Hamiltonian $\mathcal{H}_{\perp}(\tau)$. The evaluation of Eqs. \eqref{eqn:C4} -- \eqref{eqn:C6'} are straightforward. 

The $\tau$-integration and the spin $\sigma$-sum (Table \ref{tbl:spin-sum}) in Eqs. \eqref{eqn:C4}, \eqref{eqn:C5}, \eqref{eqn:C6} and \eqref{eqn:C6'} can be calculated exactly. The multi-dimensional $\tau$-integration in these equations can be done analytically by using specialized software. The exact integration yields many complicated $\beta$-dependent functions (see examples in Ref. \onlinecite{Kubo:64PrgTheoPhys(1980)}). However, in the low-temperature limit, most of these temperature-dependent terms are suppressed by the Boltzmann factor. Keeping the leading order (linear in $\beta$) and the $\mathbf{s}_{\dot{\mathbf{r}}}$-dependent terms, we obtain Eqs. \eqref{eqn:J}, \eqref{eqn:D}, and \eqref{eqn:J_inter}.

\section{\textit{ab initio} TB model}\label{app:TB}

To construct the TB Hamiltonian, we use the maximally localized Wannier functions (MLWF) to project the $d$ and $p$ orbitals onto each sublattice. The angular momentum of these orbitals is defined in the $xyz$-coordinate system (see Fig. \ref{fig:1L-CrX3}). Using this MLWF, we obtain the \textit{ab initio} Wannier Hamiltonian, $\bar{\mathcal{H}}=\bar{\mathcal{H}}_E+\bar{\mathcal{H}}_t+\bar{\mathcal{H}}^\dagger_t+\bar{\mathcal{H}}_d$, where
\begin{align}
\bar{\mathcal{H}}_E=&\sum_{\dot{\mathbf{r}}jj'}\bar{\epsilon}^m_{jj'}\bar{p}^{\dagger}_{\mathbf{r}j \sigma}\bar{p}_{\mathbf{r}j' \sigma}+\sum_{\dot{\mathbf{r}}\bar{o}\bar{o}'}\bar{\epsilon}^{\dot{m}}_{\bar{o}\bar{o}'}\bar{d}^{\dagger}_{\dot{\mathbf{r}}\bar{o} \sigma}\bar{d}_{\dot{\mathbf{r}}\bar{o}' \sigma}\label{eqn:barHE}\\
\bar{\mathcal{H}}_t=&\sum_{\dot{\mathbf{r}}\mathbf{r} }\sum_{\sigma}\bar{t}_{\mathbf{r}j ,\dot{\mathbf{r}}\bar{o}}\bar{p}^{\dagger}_{\mathbf{r}j \sigma}\bar{d}_{\dot{\mathbf{r}}\bar{o} \sigma},\label{eqn:barHt}\\
\bar{\mathcal{H}}_d=&\sum_{\dot{\mathbf{r}}\mathbf{r} }\sum_{\sigma}\bar{u}_{\dot{\mathbf{r}}'\bar{o}',\dot{\mathbf{r}}\bar{o}}\bar{d}^{\dagger}_{\dot{\mathbf{r}}'\bar{o}'\sigma}\bar{d}_{\dot{\mathbf{r}}\bar{o} \sigma},\label{eqn:barHd}
\end{align}
with $\bar{t}_{\dot{\mathbf{r}}\bar{o} \mathbf{r}j }$ and $\bar{u}_{\dot{\mathbf{r}}'\bar{o}',\dot{\mathbf{r}}\bar{o}}$ being the TB constants that are directly obtained from WANNIER90.

We note that, although the hopping Hamiltonian $\bar{\mathcal{H}}'$ is simple to construct by using these projected MLWF, this MLWF is not the basis that we use in Eqs. \eqref{eqn:HE}-\eqref{eqn:Vlambda}. As a result, the onsite Hamiltonian $\bar{\mathcal{H}}_E$ is not diagonal. Hence, in the Wannier Hamiltonian in Eqs. \eqref{eqn:barHE}, \eqref{eqn:barHt}, and \eqref{eqn:barHd} we use $\bar{o}=d_{z^2},d_{xz},d_{yz},d_{x^2-y^2},d_{xy}$ for the $d$-orbital indices [Fig.\ref{fig:1L-CrX3}(c)] and $j$ for the $p$-orbital to label this projected MLWF basis (instead of $\dot{\alpha}$ and $\alpha$). In order to connect with our microscopic model $\mathcal{H}$, we perform a unitary transformation to diagonalized $\bar{\mathcal{H}}_E$ and assume that the transformed TB constants from $\bar{\mathcal{H}}$ are approximately equal to those  in Eqs. \eqref{eqn:Ht} and \eqref{eqn:Hd}. Namely, we identify
\begin{align}
t_{\dot{\mathbf{r}}\dot{\alpha},\mathbf{r}\alpha}\approx&\sum_{\bar{o}}
\mathcal{W}^{\dot{\mathbf{r}}\dagger}_{\dot{\alpha}\bar{o}}
\bar{t}_{\dot{\mathbf{r}}\bar{o},\mathbf{r}j}\mathcal{U}^{\mathbf{r}}_{j\alpha},\label{eqn:U-trans}\\
u_{\dot{\mathbf{r}}\dot{\alpha},\dot{\mathbf{r}}'\dot{\alpha}'}\approx&\sum_{\bar{o}\bar{o}'}
\mathcal{W}^{\dot{\mathbf{r}}\dagger}_{\dot{\alpha}\bar{o}}
\bar{u}_{\dot{\mathbf{r}}\bar{o},\bar{\mathbf{r}}'\bar{o}'}\mathcal{W}^{\dot{\mathbf{r}}'}_{\dot{\alpha}'\bar{o}'},\label{eqn:W-trans}
\end{align}
where $\mathcal{U}^{\mathbf{r}}_{j\alpha}$ and $\mathcal{W}^{\dot{\mathbf{r}}}_{\dot{\alpha}\bar{o}}$ are the unitary matrices that diagonalize $\bar{\epsilon}^{\mathbf{r}}_{jj'}$ ($p$-orbital) and $\bar{\epsilon}^{\dot{\mathbf{r}}}_{\bar{o}\bar{o}'}$ ($d$-orbital) in $\bar{\mathcal{H}}_E$. Therefore, $\mathcal{U}^{m}_{j\alpha}$ is related to the rotational matrix in Eq.\eqref{eqn:R},
\begin{equation}
\mathcal{U}^{\mathbf{r}}_{j\alpha}=\mathcal{R}^{\mathbf{r}}_{j\alpha},
\end{equation}
which defines the rotation for the spin-orbit coupling Hamiltonian in \eqref{eqn:Vlambda}. The unitary matrix
$\mathcal{W}^{\dot{\mathbf{r}}\dagger}_{\dot{\alpha}\bar{o}}$ transforms the projected MLWF basis ($d_{z^2},d_{xz},d_{yz},d_{x^2-y^2},d_{xy}$) to $d$ orbitals in the trigonal basis [Fig.\ref{fig:1L-CrX3}(b)]. 

Here, we stress that the Wannier Hamiltonian $\bar{\mathcal{H}}$ in Eqs. \eqref{eqn:barHE}-\eqref{eqn:barHd} is not identical to $\mathcal{H}$, since $\bar{\mathcal{H}}$ does not have Hubbard and Hund's interacting Hamiltonian. Therefore, the direct identification of the TB constants between $\mathcal{H}$ and $\bar{\mathcal{H}}$ may misinterpret the DFT result. Also, Eqs.\eqref{eqn:barHE}-\eqref{eqn:barHd} are constructed by using the ferromagnetic ground state, since the nonmagnetic state may not yield the quasiparticle’s dispersion that is measured from the correct ground state. Because of the spin-polarized DFT calculation, the TB constants of the Wannier Hamiltonian in Eqs. \eqref{eqn:barHE}-\eqref{eqn:barHd} become spin-dependent. This poses another challenge for connecting the Wannier Hamiltonian to the spin-independent TB Hamiltonian in Eqs. \eqref{eqn:HE}-\eqref{eqn:Hd}.

The onsite energies in the Wannier Hamiltonian [diagonalized Eq. \eqref{eqn:barHE}] are spin-dependent with an evident splitting between majority spin (parallel to ground state) and minority spin (antiparallel to ground state). This splitting is the consequence of interacting effects between electrons which are automatically taken into account by DFT calculation. Therefore, the onsite energies should identify as the quasiparticle excitations of the Mott insulator (listed in Table\ref{tbl:eigenenergy}) instead of the non-interacting onsite energies in Eq. \eqref{eqn:HE}. Similar spin-dependent behavior is also found in the hopping constants of the Wannier Hamiltonian due to the electron correlations in the Kohn-Sham spectrum. This makes the identification of the TB constants of the Wannier Hamiltonian to \eqref{eqn:Ht} and \eqref{eqn:Hd} ambiguous. However, we found that the minority-spin onsite energies in the $e_g$ bands may not correspond to eigenstates in Table \ref{tbl:eigenenergy} (due to Hund interaction). Thus, the connection of the minority spin spectrum in the Wannier Hamiltonian to our model in Eqs. \eqref{eqn:HE}-\eqref{eqn:Hd} is unclear. Therefore, we use the TB constants from the majority spin Wannier Hamiltonian for Eqs. \eqref{eqn:U-trans} and \eqref{eqn:W-trans} to obtain the TB constants for Eqs. \eqref{eqn:Ht} and \eqref{eqn:Hd}. The result is summarized in Tables \ref{tbl:TBpd} and \ref{tbl:TBdd}.

Nevertheless, we also perform a similar analysis by using the minority spin TB constants from the Wannier Hamiltonian. First, we note that the difference of the TB constants between the majority and minority spin Wannier Hamiltonian is about $\pm25\%$. Despite the difference, both of the Hamiltonians preserve the same signs for all TB constants. In the minority spin calculation, we found that the estimated exchange coupling increases from 2.53meV (majority) to 3.98meV. In any case, the Wannier TB approach for estimating the exchange coupling is limited by the spin-dependent properties of its TB constants due to interacting effects. However, we argue that this approach still yields a reasonable estimate for the TB constants in Eqs. \eqref{eqn:Ht} and \eqref{eqn:Hd}.

\bibliography{CrX3}

\end{document}